\documentclass[letterpaper,12pt]{article}
\usepackage{axodraw}
\usepackage{epsf}
\usepackage{cite}
\usepackage[dvips]{graphicx}

\def\theequation{\arabic{section}.\arabic{equation}}

\begin{document}

\begin{flushright}
   MC-TH-2003-09\\[-0.15cm]
   hep-ph/0309342 \\[-0.15cm]
   October 2003
\end{flushright}

\begin{center}
{\LARGE {\bf Resonant Leptogenesis}}\\[1.5cm]
{\large Apostolos Pilaftsis and Thomas E.~J. Underwood}\\[0.3cm] 
{\em Department of Physics and Astronomy, University of Manchester,\\ 
Manchester M13 9PL, United Kingdom}
\end{center}

\vspace{1.5cm} 

\centerline{\bf ABSTRACT} 
\noindent
{\small We  study the scenario   of thermal leptogenesis in  which the
leptonic   asymmetries are resonantly  enhanced  through the mixing of
nearly degenerate heavy Majorana neutrinos  that have mass differences
comparable to their decay widths.  Field-theoretic issues arising from
the proper   subtraction of    real    intermediate states   from  the
lepton-number-violating    scattering    processes are   addressed  in
connection with an  earlier developed resummation approach to unstable
particle   mixing  in decay   amplitudes.    The  pertinent  Boltzmann
equations  are  numerically solved  after  the enhanced heavy-neutrino
self-energy  effects on  scatterings  and the  dominant gauge-mediated
collision terms are included.  We  show that resonant leptogenesis can
be realized  with heavy  ~~~Majorana neutrinos  even  as light  as  $\sim
1$~TeV, in complete accordance with the  current solar and atmospheric
neutrino data. }


\newpage

\section{Introduction}

The  recent  results from  the  Wilkinson  Microwave Anisotropy  Probe
(WMAP)  satellite  have dramatically  improved  the  accuracy of  many
cosmological  parameters~\cite{WMAP},  thus signalling  a  new era  of
precision cosmology.  For the first time, the baryon--to--photon ratio
of number  densities $\eta_B$ has  been measured to  the unprecedented
precision  of  less  than   10\%.   The~reported  value  for  $\eta_B$
is~\cite{WMAP}
\begin{equation}
  \label{BAUexp} 
\eta_B\ \equiv\ \frac{n_B}{n_\gamma}\ =\ 
                          6.1\, ^{+0.3}_{-0.2}\,\times 10^{-10}\, ,
\end{equation}
where  $n_B  =  n_b  - n_{\bar{b}}$  and   $n_\gamma$  are  the number
densities  of the net  baryon number  $B$ and photons  at the  present
epoch, respectively.

Many    theoretical    models     have    been  suggested      in  the
literature~\cite{KT,BAUreview} in order to explain the presently small
but non-zero value    of  $\eta_B$  that quantifies   the    so-called
cosmological Baryon Asymmetry in the  Universe (BAU).  One of the most
attractive  as   well as   field-theoretically  consistent scenarios of
baryogenesis is  the one proposed  by Fukugita and Yanagida~\cite{FY}.
In this  model,  out-of-equilibrium  $L$-violating decays  of  singlet
neutrinos $N_i$  with Majorana  masses  considerably larger  than  the
critical temperature   $T_c\approx 100$--200~GeV produce  initially an
excess in the lepton number $L$.  This excess in $L$ is then converted
into the   observed $B$ asymmetry through  $(B+L)$-violating sphaleron
interactions~\cite{NSM,KRS},  which  are  in thermal  equilibrium  for
temperatures ranging from $T_c$ up to $10^{12}$~GeV~\cite{AMcL,BS,HT}.
Many studies have been  devoted to analyze in  detail this scenario of
baryogenesis    through   leptogenesis~\cite{epsilonprime,MAL,Paschos,
CRV,APRD,ELN,LV,BCST,Hambye,DI,BBP,GCBetal,FHY,JCP,NT,ERY,WR,BFT,BDPS,AFS,
Anupam,CT,APreview,KH}.

In the  last few years,  the on-going neutrino experiments,  mainly at
Super-K~\cite{Super-K} and  SNO~\cite{SNO}, have been  able to address
another   important   question    in   particle   and   astro-particle
physics~\cite{ADD}.  Their analyses  have offered overwhelming support
to  the theoretical  idea that  the ordinary  neutrinos have  tiny but
non-zero  masses  and mixings~\cite{PMNS},  thereby  enabling them  to
oscillate  from one  type  of lepton  to  another~\cite{MSW}.  In  the
Standard Model~(SM) neutrinos are strictly massless.  An economical as
well as natural solution to this problem can be achieved by augmenting
the SM  field content with  right-handed (singlet) neutrinos.   By the
same token, bare Majorana masses that violate the lepton number by two
units are allowed  to be added to the Lagrangian.   The scale of these
singlet  masses is  rather model-dependent  and may  range  from about
1~TeV    in   Left-Right   Symmetric~\cite{PS,Moh/Sen}    or   certain
E${}_6$~\cite{witten}  models  up to  $10^{16}$~GeV  in typical  Grand
Unified    Theories    (GUTs)    such   as    SO(10)~\cite{FM,Wol/Wyl}
models. {}From the low-energy point of view, the large Majorana masses
present in the complete neutrino mass  matrix give rise to a kind of a
seesaw  mechanism~\cite{seesaw}, through which  the phenomenologically
favoured values for neutrino masses of order 0.1~eV and smaller can be
explained without unnaturally suppressing  the Yukawa couplings of the
theory.

One  of  the  central   questions  that  several  articles  have  been
addressing  recently  is to  which  extent  the afore-mentioned  heavy
Majorana neutrinos  can be responsible  for both the observed  BAU and
the  neutrino oscillation  data,  including possible  data from  other
non-accelerator   experiments.    In  this   context,   it  has   been
found~\cite{DI,BBP}  that  if  the  heavy singlet  neutrinos  have  an
hierarchical mass spectrum, a  lower bound of about $10^8$--$10^9$~GeV
on the leptogenesis  scale can be derived.  In  the derivation of this
lower  bound, the  size of  the leptonic  asymmetry between  the heavy
Majorana neutrino decay into a  lepton doublet $L$ and a Higgs doublet
$\Phi$,  $N_i\to L\Phi$,  and its  respective charge  and  parity (CP)
conjugate mode,  $N_i\to L^C\Phi^\dagger$, plays a key  role. In other
words,  the larger the  leptonic CP~asymmetry,  the smaller  the lower
bound on the leptogenesis scale becomes.

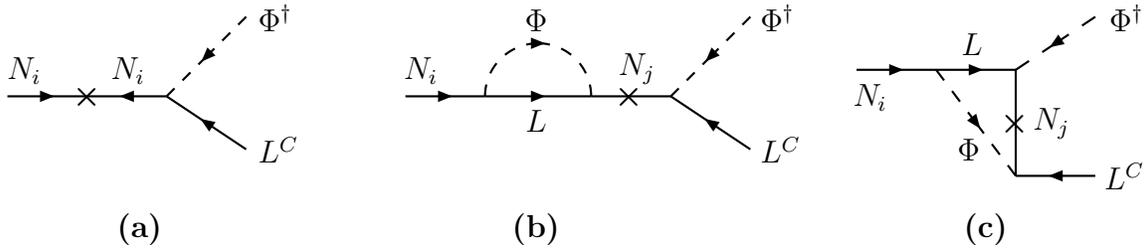
\begin{figure}

\begin{center}
\begin{picture}(450,120)(0,0)
\SetWidth{0.8}

\ArrowLine(20,60)(50,60)\ArrowLine(80,60)(50,60)
\Text(50,60)[]{{\boldmath $\times$}}
\ArrowLine(110,40)(80,60)\DashArrowLine(110,90)(80,60){5}
\Text(20,65)[bl]{$N_i$}\Text(72,65)[br]{$N_i$}
\Text(115,40)[l]{$L^C$}\Text(115,90)[l]{$\Phi^\dagger$}

\Text(70,10)[]{\bf (a)}

\ArrowLine(170,60)(200,60)\ArrowLine(200,60)(240,60)
\Line(240,60)(270,60)\Text(255,60)[]{{\boldmath $\times$}}
\DashArrowArcn(220,60)(20,180,0){5}
\ArrowLine(300,40)(270,60)\DashArrowLine(300,90)(270,60){5}
\Text(170,65)[bl]{$N_i$}\Text(265,65)[br]{$N_j$}
\Text(220,85)[b]{$\Phi$}\Text(220,55)[t]{$L$}
\Text(305,40)[l]{$L^C$}\Text(305,90)[l]{$\Phi^\dagger$}

\Text(220,10)[]{\bf (b)}

\ArrowLine(340,70)(370,70)\ArrowLine(370,70)(400,70)
\Line(400,70)(400,30)\ArrowLine(430,30)(400,30)
\Text(401,50)[]{{\boldmath $\times$}}  
\DashArrowLine(430,90)(400,70){5}\DashArrowLine(370,70)(400,30){5}
\Text(340,65)[lt]{$N_i$}\Text(385,80)[]{$L$}\Text(408,50)[l]{$N_j$}
\Text(437,90)[l]{$\Phi^\dagger$}\Text(435,30)[l]{$L^C$}
\Text(388,40)[r]{$\Phi$}

\Text(390,10)[]{\bf (c)}

\end{picture}
\end{center}

\caption{\em Feynman diagrams contributing to the $L$-violating decays
of heavy Majorana neutrinos, $N_i \to L^C \Phi^\dagger$, where $L$ and
$\Phi$ represent  lepton and  Higgs-boson  iso-doublets, respectively:
(a)  tree-level graph, and  one-loop   (b) self-energy and (c)  vertex
graphs.}\label{f1}

\end{figure}

As is shown in Fig.~\ref{f1} in a Feynman--diagrammatic way, there are
two  one-loop   graphs that  contribute to   the CP-violating leptonic
asymmetry.  In particular,   the interference of the tree-level  decay
amplitude  with the absorptive parts of   the one-loop self-energy and
vertex graphs  violates  CP and hence  gives  rise to a  non-vanishing
leptonic  asymmetry.   These self-energy and vertex  contributions are
often termed  in the literature~\cite{IKS,APreview} $\varepsilon$- and
$\varepsilon'$-types  of     CP  violation,    respectively.    Unlike
$\varepsilon'$-type~\cite{FY,epsilonprime,MAL},  $\varepsilon$-type CP
violation can be considerably enhanced~\cite{Paschos,CRV,APRD} through
the mixing of two nearly degenerate heavy Majorana neutrinos.

The  fact  that $\varepsilon$-type  CP  violation  can become  several
orders of magnitude larger than $\varepsilon'$-type CP violation might
raise  concerns  on  the  validity of  perturbation  theory.   Indeed,
finite-order  perturbation theory  breaks down  if two  heavy Majorana
neutrinos  become  degenerate.  However,  based  on a  field-theoretic
approach    that   consistently    resums    all   the    higher-order
self-energy-enhanced diagrams,  it has been  shown in~\cite{APRD} that
the leptonic  CP asymmetry is not only  analytically well-behaved, but
it  can also be  of order  {\em unity}  if two  of the  heavy Majorana
neutrinos  have mass  differences  comparable to  their decay  widths.
Because of  this resonant enhancement of the  leptonic asymmetries, we
call this scenario of leptogenesis {\em resonant leptogenesis}.

An immediate consequence of  resonant leptogenesis is that the singlet
mass     scale     can    be     drastically     lowered    to     TeV
energies~\cite{APRD,APreview}.   However, these previous  studies have
not  considered possible  limits  that may  arise  from the  presently
better  constrained  light-neutrino sector.   In  this  paper we  will
analyze the scenario of resonant  leptogenesis in light of the current
solar   and    atmospheric   neutrino   data~\cite{nudata,JWFV}.    In
particular,  we will show  that resonant  leptogenesis can  occur with
heavy Majorana  neutrinos even  as light as  $\sim 1$~TeV,  within the
framework  of light-neutrino  scenarios with  normal or  inverted mass
hierarchy   and  large   $\nu_e$-$\nu_\mu$   and  $\nu_\mu$-$\nu_\tau$
mixings,  namely  within   schemes  currently  suggested  by  neutrino
oscillation data.

Our predictions  for the BAU are  obtained after numerically solving a
network of Boltzmann Equations (BEs) related  to leptogenesis.  In our
analysis, we include the dominant collision terms  that account for $2
\to 2$ scatterings involving  the SU(2)$_L$ and U(1)$_Y$ gauge bosons.
Furthermore,   the  resonantly   enhanced   CP-violating  as  well  as
CP-conserving effects  on   the   scattering   processes   thanks   to
heavy-neutrino mixing are  taken  into account.  To  the best  of  our
knowledge, these two important contributions  to the BEs have not been
considered in the existing literature before.

The proper description of the dynamics of unstable particles and their
mixing  phenomena is  a subtle  issue within  the context  of  a field
theory.   To deal  with  this problem,  one  is compelled  to rely  on
resummation  approaches   to  unstable  particles   that  consistently
maintain   all   desirable   field-theoretic   properties,   such   as
gauge-invariance, analyticity and unitarity~\cite{PP,Stuart}.  In this
context, a  resummation approach to unstable particle  mixing in decay
amplitudes was developed in~\cite{APRD} which preserves CPT invariance
and unitarity~\cite{AP,APreview}.

In this paper we address another important issue related to the proper
subtraction  of the so-called real intermediate  states (RIS) from the
$L$-violating $2\to 2$ scattering   processes  that result  from   the
exchange of unstable particles in the $s$-channel.  Such a subtraction
is necessary  in order to  avoid double-counting in the BE's~\cite{KW}
from  the  already considered  $1\to  2$ decays  and $2\to  1$ inverse
decays of the unstable  particles, namely those associated  with heavy
Majorana neutrino decays.  By examining the analytic properties of the
pole    and    residue  structures~\cite{Stuart,AP}   of  a   resonant
$L$-violating scattering amplitude, we  can  identify the part of  the
$2\to 2$ amplitude that contains RIS contributions only.  We find that
the so-derived resonant amplitude exhibits the very same analytic form
with the one   obtained    with  an  earlier   proposed    resummation
method~\cite{APRD}.  Since the present  derivation  does not  rely  on
resorting to a  kind of Lehmann--Symanzik--Zimmermann  (LSZ) reduction
formalism~\cite{LSZ}  for the  decaying  unstable particle, it  offers
therefore  a firm  and  independent support to   the earlier treatment
presented in~\cite{APRD}.

The   paper is organized   as  follows: in Section~\ref{sec:models} we
discuss the generic structure of a  heavy Majorana-neutrino model that
possesses a low  singlet  scale and  predicts nearly degenerate  heavy
Majorana    neutrinos.    Employing      the    Froggatt--Nielsen~(FN)
mechanism~\cite{FN}, we also  put  forward a generic texture   for the
light  neutrino mass matrices  that enable an  adequate description of
the    present    solar   and    atmospheric    neutrino    data.   In
Section~\ref{sec:RIS} we  address   field-theoretic issues that  arise
from  the proper subtraction  of RIS from the $L$-violating scattering
processes.  In particular, we  explicitly demonstrate how the resonant
part of the scattering amplitude is intimately related to the resummed
decay amplitude  derived earlier by  means  of an LSZ-type resummation
approach~\cite{APRD}. Analytic formulae related to the general case of
three     heavy-Majorana-neutrino     mixing   are      given       in
Appendix~\ref{app:mixing}.  In  Section~\ref{sec:BEs} we  derive   the
relevant network    of   BE's for resonant    leptogenesis,  where the
gauge-mediated    collision   terms   and    the   resonantly enhanced
CP-violating as well as CP-conserving contributions to scatterings due
to heavy neutrino mixing are taken into account.  Analytic expressions
of  reduced   cross-sections  for  all relevant   $2\to  2$ scattering
reactions are presented in Appendix~\ref{app:CT}.  Our conclusions are
summarized in Section~\ref{sec:concls}.

\setcounter{equation}{0}
\section{Low-Scale Heavy Majorana-Neutrino Model\\ and
        Neutrino Data}\label{sec:models}

In this section, we first set up  our conventions by briefly reviewing
the low-energy   structure of a  minimally  extended  SM that includes
heavy Majorana neutrinos. We then put  forward a generic scenario that
predicts nearly degenerate heavy  Majorana neutrinos at the  TeV scale
and can naturally be realized  by means of the FN~mechanism~\cite{FN}.
In this generic scenario, the  light-neutrino sector admits the  Large
Mixing  Angle (LMA) Mikheyev--Smirnov--Wolfenstein    (MSW)~\cite{MSW}
solution and so may  explain the solar  neutrino data through a  large
$\nu_e$-$\nu_\mu$ mixing.  The light-neutrino sector also allows for a
large $\nu_\mu$-$\nu_\tau$-mixing to   account for   the   atmospheric
neutrino anomaly.  Another property of our generic scenario is that it
leads to a   mass spectrum   for the   light  neutrinos, denoted    as
$\nu_{1,2,3}$ (with  the mass convention  $m_{\nu_1} \le m_{\nu_2} \le
m_{\nu_3}$),  with normal or  inverted hierarchy, depending on whether
the lightest physical neutrino $\nu_1$ has  predominantly a $\nu_e$ or
a  $\nu_\tau$  component.   In particular,  the  generic  scenario can
accommodate  the   phenomenologically       favoured     neutrino-mass
differences~\cite{nudata,JWFV}:
\begin{equation}
  \label{nudata}
1.4\times
10^{-3}\: <\: \Delta m^2_{\rm atm}~[{\rm eV}^2]\ <\ 3.7\times
10^{-3}\,,\qquad
5.4\times
10^{-5}\: <\: \Delta m^2_\odot~[{\rm eV}^2]\ <\ 9.5\times 10^{-5}\,,
\end{equation}
at  the  3$\sigma$ confidence  level,  with  $\Delta  m^2_{\rm atm}  =
m^2_{\nu_3}  -  m^2_{\nu_2}$ and  $\Delta  m^2_\odot  = m^2_{\nu_2}  -
m^2_{\nu_1}$.

A minimal,  symmetric realization   of a  model  with heavy   Majorana
neutrinos can be   obtained by adding   to  the SM field  content  one
right-handed (singlet)  neutrino per family  $\nu_{iR}$, with $i=1, 2,
3$. The leptonic sector of this minimal model consists of the fields:
\begin{equation}
L_l\ =\   \left( \begin{array}{c} \nu_{lL}  \\  l_{lL} \end{array} \right)\
,\qquad l_{lR}\ , \qquad \nu_{iR}\ ,
\end{equation}
where  the obvious  labelling, $l=  (1,2,3) =  (e,\mu,\tau)$,  will be
employed.  At  temperatures $T$  larger than the  critical temperature
$T_c$   associated  with   the  electroweak   phase   transition,  the
$T$-dependent vacuum  expectation value (VEV)  $v(T)$ of the  SM Higgs
doublet   $\Phi$  vanishes,   i.e.\  $\langle   \Phi  (T)   \rangle  =
v(T)/\sqrt{2}  = 0$.   This is  the  epoch where  a possible  leptonic
asymmetry created by out-of-equilibrium heavy Majorana-neutrino decays
can  be actively  reprocessed into  the BAU  through  the equilibrated
$(B+L)$-violating sphaleron interactions.

At  this  epoch relevant to  leptogenesis, the  dynamics of  the early
Universe   is usually  described  by  a   Lagrangian  in  the unbroken
gauge-symmetric  phase  of the  theory.  In  this unbroken  phase, the
Lagrangian  of the  leptonic   sector of the   model under  study  may
conveniently be expressed as
\begin{equation}
  \label{Llept} {\cal L}_{\rm lept}\ =\ {\cal L}_{\rm kin}\: +\: {\cal
L}_{\rm Y}\: +\: {\cal L}_{\rm M}\, ,
\end{equation}
with 
\begin{eqnarray}
  \label{Lkin}
{\cal L}_{\rm kin} & = & \sum\limits_{i=1}^{3}\,\bigg(\,
\bar{L}_i\, i\!\not\!\partial\, L_i \: +\: 
\bar{\nu}_{iR}\, i\!\not\!\partial\, \nu_{iR}\: +\: 
\bar{l}_{iR}\, i\!\not\!\partial\, l_{iR}\,\bigg)\,,\\
  \label{LYint}
{\cal L}_{\rm Y} & = & -\, \sum\limits_{i,j=1}^{3}\, \bigg(\,
h^{\nu_R}_{ij}\, \bar{L}_i\, \tilde{\Phi}\, \nu_{jR}\: +\: 
h^l_{ij}\, \bar{L}_i\, \Phi\, l_{jR}\ +\ \mbox{H.c.}\,\bigg)\,,\\
  \label{Majmass}
{\cal  L}_{\rm  M} &=&  -\, \frac{1}{2}\, \sum\limits_{i,j=1}^{3}\, 
\bigg(\, (\bar{\nu}_{iR})^C\, (M_S)_{ij}\, \nu_{jR}\ +\ 
\bar{\nu}_{iR}\, (M_S)^*_{ij}\, (\nu_{jR})^C\, \bigg)\, .
\end{eqnarray}
In the   above, ${\cal L}_{\rm   kin}$, ${\cal L}_{\rm Y}$  and ${\cal
L}_{\rm   M}$ describe the kinetic  terms,  the Yukawa  sector and the
Majorana    masses    of  the  model,   respectively.    In  addition,
$\tilde{\Phi}=i\tau_2\Phi^*$  is the  isospin  conjugate of the  Higgs
doublet $\Phi$,  where $\tau_2$ is   the usual  Pauli matrix, and  the
superscript $C$ denotes charge conjugation.

In the  unbroken phase of  the theory, only  the singlet neutrinos are
massive. Their physical masses   can  be found by  diagonalizing   the
3-by-3  singlet Majorana   mass matrix $M_S$  in~(\ref{Majmass}).  The
matrix  $M_S$  is symmetric  and    in general complex,   and can   be
diagonalized by means of a unitary transformation
\begin{equation}
  \label{Utrans}
U^T\, M_S\, U\ =\ \widehat{M}_S\ \equiv\ 
         {\rm diag}\, \big(\,m_{N_1},\, m_{N_2},\, m_{N_3}\,\big)\,,
\end{equation}
where  $U$   is  a    $3\times  3$-dimensional unitary     matrix  and
$m_{N_{1,2,3}}$  denote the  3 physical masses  of  the heavy Majorana
neutrinos    $N_{1,2,3}$,    ordered as   $m_{N_1}  \le    m_{N_2} \le
m_{N_3}$.  Correspondingly,   the  flavour    states   $\nu_{iR}$  and
$(\nu_{iR})^C$ are related to the mass eigenstates $N_i$ through
\begin{equation}
 \label{nuRs}
\nu_{iR}\ =\  P_R\,\sum\limits_{j=1}^{3}\, U_{ij} N_j\,,\qquad
(\nu_{iR})^C\ =\  P_L\,\sum\limits_{j=1}^{3}\, U^*_{ij} N_j\,,
\end{equation}
where  $P_R = (1  + \gamma_5)/2$ and $P_L  = (1  - \gamma_5)/2$.  Note
that $\nu_{iR}$ and  $(\nu_{iR})^C$ do not  transform independently of
one  another under  a  unitary rotation.  In  the physical basis,  the
Yukawa leptonic sector reads
\begin{equation}
  \label{LYphys}
{\cal L}_{\rm Y} \ =\ -\, \sum\limits_{i,j=1}^{3}\, \bigg(\,
h^\nu_{ij}\, \bar{L}_i\, \tilde{\Phi}\, P_R\, N_j\: +\: 
\hat{h}^l_{ii}\, \bar{L}_i\, \Phi\, P_R\, l_i\quad + \quad 
\mbox{H.c.}\,\bigg)\,,
\end{equation}
where a four-component chiral  representation for all fermionic fields
should  be  understood.    In~(\ref{LYphys}),  $\hat{h}^l_{ii}$  is  a
diagonal   positive   matrix    and   $h^\nu_{ij}$   is   related   to
$h^{\nu_R}_{ij}$  through a  bi-unitary  transformation: $h^\nu\,  =\,
V_L^\dagger\, h^{\nu_R}\,  U$, where $V_L$ is a  3-by-3 unitary matrix
that transforms the left-handed charged leptons to their corresponding
mass eigenstates.   Our computations  of the leptonic  asymmetries and
collision  terms  relevant  to  leptogenesis  will  be  based  on  the
Lagrangian~(\ref{LYphys}).

Having set  the stage, it is  now instructive  to discuss the possible
flavour structure of low  singlet-scale models with  nearly degenerate
heavy Majorana neutrinos. Such a class of models may be constructed by
assuming   that lepton-number  violation  (and possibly  baryon-number
violation) occurs at very high energies at the  GUT scale $M_{\rm GUT}
\sim 10^{16}$--$10^{17}$~GeV, or even higher close to the Planck scale
$M_{\rm    Planck}     \sim   10^{19}$~GeV    through    gravitational
interactions. On the other hand, operators that conserve lepton number
are allowed to be at the TeV scale. 

Since our interest   is    to resonant  leptogenesis,  the   following
sufficient  and necessary conditions  under which leptonic asymmetries
of order unity can take place have to be satisfied  by the model under
discussion~\cite{APRD}:
\begin{equation}
  \label{CPres}
m_{N_i} \, -\, m_{N_j}\ \sim\  \frac{\Gamma_{N_{i,j}}}{2}\ ,\qquad
\frac{\big| {\rm Im}\, \big( h^{\nu\dagger} h^\nu
\big)^2_{ij}\,\big|}{
(h^{\nu\dagger} h^\nu)_{ii}\, (h^{\nu\dagger} h^\nu)_{jj}}\ \sim \ 1\, ,
\end{equation}
for a pair of heavy Majorana neutrinos $N_{i,j}$.  In (\ref{CPres}),
$\Gamma_{N_i}$ are the $N_i$ decay widths, which at the tree level are
given by
\begin{equation}
  \label{GNi}
\Gamma^{(0)}_{N_i}\ =\ \frac{(h^{\nu\dagger} h^\nu)_{ii}}{8\pi}\ m_{N_i}\; .
\end{equation}

In the following, we present  a rather generic scenario that minimally
realizes  the above requirements and still   has sufficient freedom to
describe the neutrino  data.  Our generic scenario  is based on the FN
mechanism~\cite{FN}.  Specifically,  we    introduce  two FN   fields,
$\Sigma$   and  $\overline{\Sigma}$,  with   opposite  U(1)$_{\rm FN}$
charges,    i.e.\ $Q_{\rm  FN}    (  \Sigma  )    =  -  Q_{\rm FN}   (
\overline{\Sigma} ) =  +  1$.  Under  U(1)$_{\rm FN}$, the   following
charges for the right-handed neutrinos are assigned:
\begin{equation}
  \label{QFN}
Q_{\rm  FN}\, (\nu_{1R}) \ =\ -1\,,\qquad 
Q_{\rm  FN}\, (\nu_{2R} )\ =\ +1\,,\qquad 
Q_{\rm  FN}\, (\nu_{3R} )\ =\ 0\; .
\end{equation}
In addition, all other fields, including charged leptons, are singlets
under U(1)$_{\rm FN}$. Then, the singlet mass matrix $M_S$ assumes the
generic form:
\begin{equation}
  \label{MS}
M_S\ \sim \ M\, \left(\! \begin{array}{ccc}
  \varepsilon^2 & 1 & \varepsilon\\
  1 & \bar{\varepsilon}^2 & \bar{\varepsilon}\\
  \varepsilon & \bar{\varepsilon} & M_X/M \end{array}\! \right)\,,
\end{equation}
where   $\varepsilon  = \langle    \Sigma  \rangle  /M_{\rm  GUT}$ and
$\bar{\varepsilon } =   \langle  \overline{\Sigma }  \rangle  / M_{\rm
GUT}$.  In (\ref{MS}), $M$ sets up the scale  of the leptonic symmetry
$L_e -  L_\mu$,\footnote{Similar  textures  of $M_S$  may  result from
E$_6$  theories~\cite{witten},    where  the   lepton    numbers   are
approximately broken~\cite{BGL}.}  while $M_X$ represents the scale of
$L_\tau$ violation.   It is conceivable  that these two  scales may be
different from one  another.  For the  case of our  interest, it is $M
\sim 1$~TeV, while $M_X$ is considered to be  many orders of magnitude
larger close to $M_{\rm GUT}$.

The FN mechanism also determines the strength of the Yukawa couplings.
After    spontaneous   symmetry     breaking~(SSB),     the  resulting
Dirac-neutrino mass matrix $m_D$ has the generic form
\begin{equation}
  \label{mD} 
m_D\ \equiv\ \frac{v}{\sqrt{2}}\, h\ \sim \
\frac{v}{\sqrt{2}}\, \left(\! \begin{array}{ccc} \varepsilon &
\bar{\varepsilon} & 1\\ \varepsilon & \bar{\varepsilon} & 1\\
\varepsilon & \bar{\varepsilon} & 1 \end{array}\! \right)\, ,
\end{equation}
where  $h$  is  a $3\times  3$ matrix  containing  the neutrino Yukawa
couplings,  expressed in  the   positive and  diagonal basis  of   the
respective charged-lepton Yukawa couplings.  

If one   assumes     that  $\langle  \Sigma  \rangle     \sim  \langle
\overline{\Sigma } \rangle \sim  \sqrt{M\, M_{\rm GUT}}$ and $M_X \sim
M_{\rm GUT}$, a rather simple pattern  for the mass matrices $m_D$ and
$M_S$ emerges.   In   this case, the  mass  spectrum  of   the generic
scenario  under  investigation    contains one  super-heavy    Majorana
neutrino,  with a mass $m_{N_3}  \sim M_X  \sim  M_{\rm GUT}$, and two
nearly  degenerate    heavy  Majorana   neutrinos    $N_{1,2}$,   with
$m_{N_{1,2}} \sim M$  and a  mass difference  $m_{N_1} -  m_{N_2} \sim
\varepsilon^2  M     \sim   M^2/M_{\rm    GUT}$.    Since     it    is
$\Gamma^{(0)}_{N_{1,2}} \sim \varepsilon^2 M \sim M^2/M_{\rm GUT}$, it
can  be readily seen that one  of  the crucial conditions for resonant
leptogenesis   in~(\ref{CPres}),  i.e.\    $m_{N_1}  -  m_{N_2}   \sim
\frac{1}{2}\,\Gamma_{N_{1,2}}$, can  naturally be satisfied within our
generic framework.

In the above exercise,  one should bear in mind  that the FN mechanism
can only give rise to an order-of-magnitude  estimate of the different
entries  in the  mass matrices $m_D$  and  $M_S$.  Moreover, since our
focus will be on the neutrino sector of this minimal model of resonant
leptogenesis, we  will not attempt  to explain the complete quark- and
charged-lepton-mass  spectrum   of the SM    by analyzing all possible
solutions through the FN mechanism. Such  an extensive study is beyond
the scope of the present article and may be given elsewhere.

We will  now  explicitly  demonstrate that the  mass   textures stated
in~(\ref{MS})   and~(\ref{mD})   can lead   to  viable  light-neutrino
scenarios, when  the latter are confronted  with the present solar and
atmospheric neutrino  data.   To further simplify  our  discussion, we
assume that  the  super-heavy neutrino decouples  completely  from the
light-neutrino  spectrum.  As  a result,  to  leading order  in the FN
parameters  $\varepsilon$   and  $\bar{\varepsilon}$,    the    3-by-3
light-neutrino mass-matrix ${\bf m}^\nu$ may be cast into the form:
\begin{eqnarray}
  \label{mnulight}
{\bf m}^\nu\ \approx \ -\, \frac{v^2}{2M}\, \left(\! \begin{array}{ccc}
2h_{11}h_{12} & h_{11}h_{22} + h_{12}h_{21} & h_{11}h_{32} + h_{31}h_{12} \\ 
h_{11}h_{22} + h_{12}h_{21} & 2h_{21}h_{22} & h_{21}h_{32} + h_{31}h_{22} \\
h_{11}h_{32} + h_{31}h_{12} & h_{21}h_{32} + h_{31}h_{22} & 2h_{31}h_{32}
\end{array}\!\right)\, .
\end{eqnarray}
Here, $h_{ij}$ are  the  neutrino Yukawa couplings  in the  weak basis
described    after  (\ref{mD}).  Note that effects    due  to the mass
degeneracy of the heavy Majorana  neutrinos contribute terms ${\cal O}
(\varepsilon^3 \bar{\varepsilon}, \varepsilon \bar{\varepsilon}^3)$ to
${\bf       m}^\nu$.       As       long      as      $\varepsilon\,,\
\bar{\varepsilon}\stackrel{<}{{}_\sim}  10^{-3}$,  these   sub-leading
terms do  not affect the light-neutrino  mass spectrum  and hence they
can  be   safely  neglected.   The scenarios  which   we  will address
numerically  in   Section 4.3 are   compatible with   these  limits on
$\varepsilon$ and $\bar{\varepsilon}$.

Let us now present a concrete example by considering the following set
of  Yukawa   couplings    (given   in   units   of  $\varepsilon     =
\bar{\varepsilon}$):
\begin{equation}
  \label{hmodel}
h_{11}\ =\ -\,\frac{1}{3}\,;\quad h_{12}\ =\ \frac{2}{3}\,;\quad
h_{21}\ =\ 2\,;\quad h_{22}\ =\ 1\,;\quad h_{31}\ =\ 1\,; 
\quad h_{32}\ =\ 2\,.
\end{equation}
For our  illustrations, we   also  neglect the  existence  of possible
CP-odd phases  in the Yukawa couplings.  Then, the light-neutrino mass
matrix exhibits the structure
\begin{eqnarray}
  \label{mnuappr}
{\bf m}^\nu \ \approx\ -\, \frac{v^2 \varepsilon\bar{\varepsilon} }{2M}\
\left(\! \begin{array}{ccc}
-4/9   & 1 & 0 \\ 
  1    & 4 & 5 \\
  0    & 5 & 4 \end{array}\!\right)\, .
\end{eqnarray}
It is not  difficult to see that  the above light-neutrino mass matrix
${\bf m}^\nu$ can be  diagonalized  by large $\nu_\mu$-$\nu_\tau$  and
$\nu_e$-$\nu_\mu$ mixing   angles,   i.e.\ $|\theta_{\nu_\mu\nu_\tau}|
\sim  \pi/4$ and  $|\theta_{\nu_e\nu_\mu}| \sim  \pi/6$.  Instead, the
$\nu_e$-$\nu_\tau$   mixing angle  is   estimated  to be small,  i.e.~
$|\theta_{\nu_e \nu_\tau}| \stackrel{<}{{}_\sim} 0.1$, as is suggested
by the CHOOZ  experiment~\cite{CHOOZ,JWFV}.  Furthermore, the physical
light-neutrino  masses derived  from  ${\bf m}^\nu$ are  approximately
given by
\begin{equation}
  \label{mnuth}
\big(\, m_{\nu_1},\ m_{\nu_2},\ m_{\nu_3}\,\big)\ \approx\ 
\frac{v^2 \varepsilon\bar{\varepsilon}}{2M}\ \big( 0.04,\ 1.5,\ 9 \big)\ \sim\ 
\frac{m^2_t}{M_{\rm GUT}}\ \big( 0.04,\ 1.5,\ 9 \big)\, .
\end{equation}
In deriving the last step of (\ref{mnuth}), we have used the fact that
$|\varepsilon \bar{\varepsilon}| \sim M/M_{\rm  GUT}$ and $m_t \approx
v/\sqrt{2}$.   Observe that  although   our  approach  here has   been
different, the light-neutrino masses  in (\ref{mnuth}) still  obey the
known seesaw mass relation~\cite{seesaw},  and scale independently  of
$M$.    In particular,  one  can easily   check that (\ref{mnuth})  is
compatible with the   observed light-neutrino mass differences  stated
in~(\ref{nudata}).  Even   though   the present  example  realizes   a
light-neutrino  mass spectrum   with   normal hierarchy, an   inverted
hierarchy  can easily  be obtained  by  appropriately rearranging  the
Yukawa couplings in~(\ref{hmodel}).

Our numerical estimates in Section~\ref{sec:BEs} will rely on neutrino
models that make use of the generic  structures for the matrices $M_S$
and $m_D$, given in (\ref{MS}) and (\ref{mD}), respectively.

\setcounter{equation}{0}
\section{Subtraction of RIS and Leptonic Asymmetries}\label{sec:RIS}

In this section  we wish to address an important  issue related to the
proper subtraction  of the  so-called real intermediate  states (RIS),
e.g.~heavy Majorana  neutrinos $N_i$, from the  $L$-violating $2\to 2$
scattering processes.  As we will see in Section \ref{sec:BEs}, such a
subtraction  is necessary  in order  to avoid  double-counting  in the
BEs~\cite{KW} from the already considered $1\to 2$ decays and $2\to 1$
inverse decays  of the unstable heavy Majorana  neutrinos. By studying
the analytic  properties of the pole  and the residue  structures of a
resonant $L$-violating  scattering amplitude, we are  able to identify
the  resonant  part  of  a   $2\to  2$  amplitude  that  contains  RIS
contributions  only. The  so-derived  resonant amplitude  can then  be
shown to  exhibit the  very same analytic  form with the  one obtained
with  an  earlier  proposed resummation  method~\cite{APRD}.   Another
important result of our considerations  is that we can define one-loop
resummed effective Yukawa couplings  that capture all dominant effects
of heavy Majorana-neutrino mixing and CP violation.

\subsection{Approach to the Subtraction of RIS}

Let us first consider the  simple scattering process $L\Phi \to N^*\to
L^C\Phi^\dagger$, mediated  by a single  heavy-neutrino exchange  $N$.
This exercise will help us to demonstrate  our approach to subtracting
the RIS  part of an  amplitude.  The more  realistic  case of resonant
leptogenesis with two heavy Majorana neutrinos will be discussed later
on.  To  keep things at an intuitive  level, we assume throughout this
section that all particles involved in this  process are scalar, e.g.\
scalar neutrinos or sneutrinos $\widetilde{N}_i$ that are predicted in
supersymmetric  theories~\cite{soft}.   Nevertheless,  we will discuss
the complications   that may arise   in our   considerations from  the
spinorial  nature  of the  lepton and   heavy   neutrino fields.   The
$s$-channel contribution to  the scattering amplitude ${\cal  T}(L\Phi
\to L^C\Phi^\dagger)$ due to a single $\widetilde{N}$-exchange reads:
\begin{equation}
  \label{Tsimple}
{\cal T}_s (L\Phi \to L^C\Phi^\dagger)\ =\ 
{\cal T}_A(L\Phi \to {\widetilde{N}}^*)\ 
\frac{1}{s - m^2_{\widetilde{N}} + i\,{\rm
    Im}\,\Pi_{\widetilde{N} \widetilde{N}}(s)}\
{\cal T}_B(\widetilde{N}^* \to L^C\Phi^\dagger)\ ,
\end{equation}
where the  Breit--Wigner-like propagator has been  obtained by summing
up   an    infinite   series   of    heavy   sneutrino   self-energies
$\Pi_{\widetilde{N}\widetilde{N}}  (s)$.  The  dispersive part  of the
self-energy  ${\rm  Re}\,\Pi_{\widetilde{N}\widetilde{N}} (s)$,  which
has been  omitted here,  can be suppressed  by renormalization  at the
resonant   region  $s\approx   m^2_{\widetilde{N}}$   (see  also   our
discussion    below).    Instead,    its    absorptive   part    ${\rm
Im}\,\Pi_{\widetilde{N}\widetilde{N}}      (m^2_{\widetilde{N}})     =
m_{\widetilde{N}}\Gamma_{\widetilde{N}}$  is  essential  to obtain  an
analytically  well-behaved  amplitude  at $s  =  m^2_{\widetilde{N}}$,
where $\Gamma_{\widetilde{N}}$  is the total decay width  of the heavy
sneutrino $\widetilde{N}$.

As   we  will   see  in   Section~\ref{sec:BEs}.1,  out-of-equilibrium
constraints     on    the    heavy-(s)neutrino     width    $\Gamma_N$
($\Gamma_{\widetilde{N}}$)      imply      $\Gamma_N     \ll      m_N$
($\Gamma_{\widetilde{N}}  \ll m_{\widetilde{N}}$).  In  this kinematic
regime,  the so-called  pole-dominance  or narrow-width  approximation
constitutes a very accurate approach to subtract the RIS part from the
squared matrix  element $|{\cal T}_s  (L\Phi \to L^C\Phi^\dagger)|^2$.
According to the pole-dominance approximation, we have
\begin{eqnarray}
  \label{Approx}
\frac{1}{(s - m^2_{\widetilde{N}} )^2\: +\: 
m^2_{\widetilde{N}} \Gamma^2_{\widetilde{N}} } & = & \nonumber\\
&&\hspace{-3cm}
\frac{1}{(s - m^2_{\widetilde{N}} + i m_{\widetilde{N}}
\Gamma_{\widetilde{N}} ) \,
( s - m^2_{\widetilde{N}} - im_{\widetilde{N}}\Gamma_{\widetilde{N}}) }\
\to\ \frac{\pi}{m_{\widetilde{N}}
\Gamma_{\widetilde{N}} }\ \delta (s - m^2_{\widetilde{N}})\:
\theta (\sqrt{s})\; ,\qquad
\end{eqnarray}
where  $\delta(x)$ and  $\theta  (x)$  are the  usual  Dirac and  step
functions,  respectively.   Notice  that  in~(\ref{Approx})  only  one
residue related to  the physical pole at $s  = m^2_{\widetilde{N}} - i
m_{\widetilde{N}}  \Gamma_{\widetilde{N}}$ is  considered by  means of
the  Cauchy theorem.   Substituting~(\ref{Approx})  into $|{\cal  T}_s
(L\Phi \to  L^C\Phi^\dagger)|^2$, i.e.~after squaring~(\ref{Tsimple}),
we can uniquely isolate the RIS part for this process:
\begin{equation}
  \label{TRIS}
|{\cal T}_{\rm RIS}  (L\Phi \to L^C\Phi^\dagger)|^2\ =\ 
\frac{\pi}{m_{\widetilde{N}}\Gamma_{\widetilde{N}}}\ \delta (s
- m^2_{\widetilde{N}})\:\theta (\sqrt{s})\; 
|{\cal T}_A(L\Phi \to {\widetilde{N}})|^2\ |{\cal
  T}_B({\widetilde{N}}\to L^C\Phi^\dagger)|^2\; .  
\end{equation}
This last   result  is  fully  consistent   with  the  one   presented
in~\cite{KW}.  As we will discuss  below, however, the above  approach
of identifying the  proper RIS part  of the squared  amplitude becomes
more    involved  in  the  presence    of two  strongly-mixed unstable
particles.

The above derivation of the RIS component of the squared amplitude was
based on the assumption that the heavy neutrino ${\widetilde{N}}$ is a
scalar  particle.   The spinorial  nature  of  $N$ introduces  further
complications.     It   naively    violates   the    factorized   form
of~(\ref{TRIS}),    and    $|{\cal    T}_{\rm    RIS}    (L\Phi    \to
L^C\Phi^\dagger)|^2$  can no  longer  be  written as  a  product of  a
production and a decay squared amplitude, i.e.\ it is not proportional
to   $|{\cal  T}_A(L\Phi   \to   N)|^2$  and   to  $|{\cal   T}_B(N\to
L^C\Phi^\dagger)|^2$. Instead, we find
\begin{eqnarray}
  \label{Tspin}
|{\cal    T}_{\rm    RIS}|^2 &=& \frac{\pi}{m_N\Gamma_N}\ 
\delta (s - m^2_N)\: \theta (\sqrt{s})\nonumber\\
&&\hspace{-1cm}\times\,\sum\limits_{s_{1,2,3,4}}\
\delta_{s_1s_2}\, \delta_{s_3s_4}\: {\rm Tr}\,\bigg[\,
{\cal T}_A\, u_N (p,s_1) \bar{u}_N (p,s_2)
{\cal T}_B\:{\cal T}_B^\dagger\, u_N (p,s_3) \bar{u}_N (p,s_4)\,
{\cal T}_A^\dagger\,\bigg]\; ,\qquad
\end{eqnarray}
where  $u_N (p,s)$  is the  on-shell 4-component spinor,  with $p^2  =
m^2_N$, and the trace  is understood to act on  the spinor space.  The
RIS   squared amplitude  can    be    written  in the       factorized
form~(\ref{TRIS}), only after we  perform a Fierz rearrangement of the
spinors and  have integrated over the phase  space of the  initial and
final   states in    the calculation  of   the corresponding   reduced
cross-section      (see     also        our         discussion      in
Appendix~\ref{app:CT}). Then, after  the phase-space integrations, the
only non-vanishing    Lorentz   structure that   survives  is  of  the
parity-even form:  $a\!\not\! p +   b$,  where $a$   and $b$ are  mass
dependent constants.  Based on this observation,  it can be shown that
the  final result is fully equivalent  to (\ref{TRIS}), and amounts to
substituting into~(\ref{Tspin}):
\begin{equation}
  \label{Fierz}
\delta_{s_1 s_2}\, \delta_{s_3 s_4}\ \to \  \delta_{s_1 s_4}\,
\delta_{s_2 s_4}\; .
\end{equation}
It is important  to  note   here  that the  above spin   de-correlated
subtraction of  RIS  carrying    spin   can always be     carried  out
independently of  the  number   of  the  exchanged particles    in the
$s$-channel, such as  heavy neutrinos, provided the resonant amplitude
itself can be written  as  a sum of  single-pole resonance  terms that
have the simple factorized form of~(\ref{Tsimple}).  We will elucidate
this point below, while deriving the RIS squared  amplitude due to the
exchange of two heavy Majorana neutrinos.

\begin{figure}[t]
\begin{center}
\begin{picture}(400,100)(0,0)
\SetWidth{0.8}

\DashArrowLine(0,70)(20,50){4}\ArrowLine(0,30)(20,50)
\Text(0,75)[b]{$\Phi$}\Text(0,25)[t]{$L$}
\Line(20,50)(60,50)\GCirc(40,50){7}{0.9}
\Text(25,45)[t]{$N_1$}\Text(55,45)[t]{$N_1$}
\Text(13,50)[r]{\footnotesize $A$}\Text(67,50)[l]{\footnotesize $B$}
\DashArrowLine(80,70)(60,50){4}\ArrowLine(80,30)(60,50)
\Text(80,75)[b]{$\Phi^\dagger$}\Text(80,25)[t]{$L^C$}

\Text(40,0)[]{\bf (a)}

\DashArrowLine(110,70)(130,50){4}\ArrowLine(110,30)(130,50)
\Text(110,75)[b]{$\Phi$}\Text(110,25)[t]{$L$}
\Line(130,50)(170,50)\GCirc(150,50){7}{0.9}
\Text(135,45)[t]{$N_1$}\Text(165,45)[t]{$N_2$}
\Text(123,50)[r]{\footnotesize $A$}\Text(177,50)[l]{\footnotesize $B$}
\DashArrowLine(190,70)(170,50){4}\ArrowLine(190,30)(170,50)
\Text(190,75)[b]{$\Phi^\dagger$}\Text(190,25)[t]{$L^C$}

\Text(150,0)[]{\bf (b)}

\DashArrowLine(220,70)(240,50){4}\ArrowLine(220,30)(240,50)
\Text(220,75)[b]{$\Phi$}\Text(220,25)[t]{$L$}
\Line(240,50)(280,50)\GCirc(260,50){7}{0.9}
\Text(245,45)[t]{$N_2$}\Text(275,45)[t]{$N_1$}
\Text(233,50)[r]{\footnotesize $A$}\Text(287,50)[l]{\footnotesize $B$}
\DashArrowLine(300,70)(280,50){4}\ArrowLine(300,30)(280,50)
\Text(300,75)[b]{$\Phi^\dagger$}\Text(300,25)[t]{$L^C$}

\Text(260,0)[]{\bf (c)}

\DashArrowLine(330,70)(350,50){4}\ArrowLine(330,30)(350,50)
\Text(330,75)[b]{$\Phi$}\Text(330,25)[t]{$L$}
\Line(350,50)(390,50)\GCirc(370,50){7}{0.9}
\Text(355,45)[t]{$N_2$}\Text(385,45)[t]{$N_2$}
\Text(343,50)[r]{\footnotesize $A$}\Text(397,50)[l]{\footnotesize $B$}
\DashArrowLine(410,70)(390,50){4}\ArrowLine(410,30)(390,50)
\Text(410,75)[b]{$\Phi^\dagger$}\Text(410,25)[t]{$L^C$}

\Text(370,0)[]{\bf (d)}

\end{picture}
\end{center}
\caption{\em Resummed diagrams  contributing to  the resonant part  of
the  $2\to 2$  scattering     amplitude of  the process     $L\Phi \to
L^C\Phi^\dagger$.  Depending on the  context, $L,\ N_{1,2}$ may denote
scalar or fermion particles (see also text).}\label{fig2}
\end{figure}
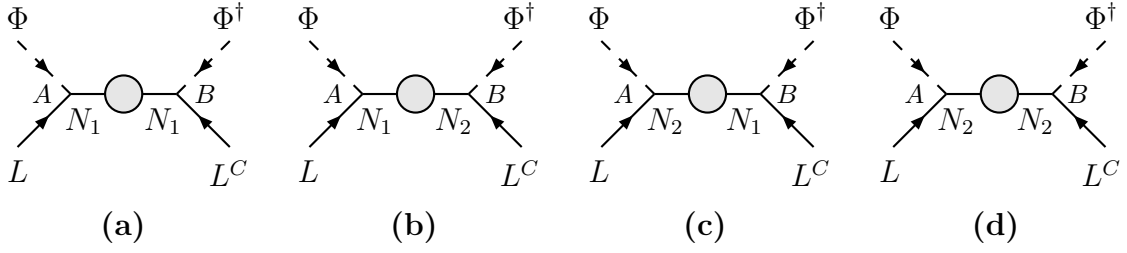

Let us therefore turn our attention to the case  of two heavy Majorana
neutrinos   $N_1$  and  $N_2$.     Analytic  expressions for  unstable
particle-mixing   effects with three  heavy   neutrinos  are given  in
Appendix~\ref{app:mixing}.  Again, we initially  assume that the heavy
neutrinos  $N_1$ and   $N_2$  are scalar  particles, i.e.\  sneutrinos
${\widetilde{N}}_{1,2}$, but  we   will discuss  in  Section 3.2   the
complications originating from their spinorial nature.  As is shown in
Fig.~\ref{fig2}, the $s$-dependent  part ${\cal T}_s$ of the amplitude
${\cal T} (L\Phi\to L^C\Phi^\dagger)$ may conveniently be expressed as
\begin{equation}
  \label{Ts}
{\cal T}_s (s)\ =\ \Gamma^A_1\, \Delta_{11}(s)\, \Gamma^B_1\:
+\: \Gamma^A_1\, \Delta_{12}(s)\, \Gamma^B_2\: +\: 
\Gamma^A_2\, \Delta_{21}(s)\, \Gamma^B_1\:
+\: \Gamma^A_2\, \Delta_{22}(s)\, \Gamma^B_2\; .
\end{equation}
In~(\ref{Ts}),  $\Gamma^{A}_{1,2}\ (\Gamma^{B}_{1,2})$  represent  the
vertices     $\Phi L    {\widetilde{N}}_{1,2}$      ($\Phi^\dagger L^C
{\widetilde{N}}_{1,2}$) that include the wave-functions of the initial
and final  states.  Analogously with   the single heavy-sneutrino case
described  above, $\Delta_{ij}(s)$   (with $i,j    =  1,2$) are    the
corresponding         ${\widetilde{N}}_i{\widetilde{N}}_j$-propagators
obtained  by  resumming    an  infinite series   of   heavy  sneutrino
self-energy graphs~\cite{APRD}:
\begin{eqnarray} 
  \label{D11}
\Delta_{11}(s) &=& \left[ \, s\, -\, m^2_{{\widetilde{N}}_1}
+\Pi_{11}(s)-\, \frac{\Pi^2_{12}(s)}{s - m^2_{{\widetilde{N}}_2} + 
\Pi_{22}(s)}\,\right]^{-1}\,,\nonumber\\
  \label{D22}
\Delta_{22}(s) &=& \left[ \, s\, -\, m^2_{{\widetilde{N}}_2}
+\Pi_{22}(s)-\, \frac{\Pi^2_{12}(s)}{s - m^2_{{\widetilde{N}}_1} + 
\Pi_{11}(s)}\,\right]^{-1}\, ,\\
  \label{D12}
\Delta_{12}(s) &=& \Delta_{21}(s)\ =\
-\, \Pi_{12}(s) \Bigg[ \Big( s - m^2_{{\widetilde{N}}_1} + \Pi_{11}(s) \Big)
\Big( s - m^2_{{\widetilde{N}}_2} + \Pi_{22}(s)\Big)\, -\, \Pi^2_{12}(s)\,
\Bigg]^{-1}\, ,\nonumber
\end{eqnarray}
where $\Pi_{12}(s) = \Pi_{21}(s)$.  We assume that the heavy sneutrino
self-energies  $\Pi_{ij} (s)$ have   already been renormalized in  the
on-shell (OS) scheme, i.e.\ they satisfy the properties:
\begin{equation}
  \label{OSren}
{\rm Re}\, \Pi_{ij} (m^2_{{\widetilde{N}}_i})\ =\
{\rm Re}\, \Pi_{ij} (m^2_{{\widetilde{N}}_j})\ =\ 0\, ,\qquad
\lim_{s\to m^2_{{\widetilde{N}}_i}}\,
\frac{{\rm Re}\, \Pi_{ii} (s) }{s -  m^2_{{\widetilde{N}}_i}} \ =\ 0\; ,
\end{equation}
where  ${\rm Re}$  indicates  that  only the  dispersive  part of  the
self-energies   must   be   considered.    Further   details   on   OS
renormalization in scalar theories may be found in~\cite{AP}.

To  technically  facilitate  our   discussion,  it  is  convenient  to
introduce the following abbreviations:
\begin{eqnarray}
  \label{auxi}
D_{ij}(s) & = & \delta_{ij}\, (s-m^2_{{\widetilde{N}}_i})\ 
                                    +\  \Pi_{ij} (s)\;,\nonumber\\
D(s) & = & D_{11} (s)\, D_{22}(s)\: - \: D_{12}(s)\,D_{21}(s)\; .
\end{eqnarray}
With       the       above       definitions,       the       resummed
${\widetilde{N}}_i{\widetilde{N}}_j$-propagators   in~(\ref{D22})  can
now be expressed in the simplified forms:
\begin{equation}
  \label{Dresum}
\Delta_{11} (s) \ =\ \frac{D_{22} (s)}{D (s)}\ ,\quad
\Delta_{22} (s) \ =\ \frac{D_{11} (s)}{D (s)}\ ,\quad
\Delta_{12} (s) \ =\ \Delta_{21} (s) \ =\ -\,\frac{D_{12} (s)}{D (s)}\
\ =\ -\,\frac{D_{21} (s)}{D (s)}\ .
\end{equation}
In addition, we introduce the quantity
\begin{equation}
  \label{WVF}
Z_i (s)\ =\  \bigg(\, \frac{d}{ds}\, \Delta^{-1}_{ii}(s)\,\bigg)^{-1}\;.
\end{equation}
In  the  OS  scheme,   with  all  contributions  from  unitarity  cuts
neglected,  we obtain  the  known  relation for  the  residues of  the
diagonal  propagators:  $Z^{\rm OS}_i  (m^2_{{\widetilde{N}}_i})~=~1$.
However, $Z_i (m^2_{{\widetilde{N}}_i})$ is in general complex, but UV
finite at order $(h^\nu)^2$.

The two complex  pole positions $s_{{\widetilde{N}}_{1,2}}$ associated
with  the heavy sneutrinos  ${\widetilde{N}}_{1,2}$ are  determined by
the equation $D(s_{{\widetilde{N}}_{1,2}}) = 0$, where $D(s)$ is given
in~(\ref{auxi}).   Since each  resummed  propagator $\Delta_{ij}  (s)$
given   in~(\ref{Dresum})  contains   two  complex   poles  at   $s  =
s_{{\widetilde{N}}_{1,2}}$,     it    can     be     expanded    about
$s_{{\widetilde{N}}_{1,2}}$ as follows:
\begin{eqnarray}
  \label{D11exp}
\Delta_{11}(s) &=& \frac{D_{22}(s)}{D(s)}\,\bigg|_{s\approx
  s_{{\widetilde{N}}_1}}\  
+\quad \frac{D_{22}(s)}{D(s)}\,\bigg|_{s\approx
  s_{{\widetilde{N}}_2}}\ +\ \dots\nonumber\\ 
&=&  \frac{Z_1(s)}{s - s_{{\widetilde{N}}_1}}\ +\
\frac{D_{12}(s)}{D_{11}(s)}\,\frac{D_{11}(s)}{D(s)}\,
\frac{D_{12}(s)}{D_{11}(s)}\,\bigg|_{s\approx s_{{\widetilde{N}}_2}}\
+\ \dots\nonumber\\ 
&=&  \frac{Z_1(s)}{s - s_{{\widetilde{N}}_1}}\ +\
\frac{D_{12}(s)}{D_{11}(s)}\,\frac{Z_2(s)}{s - s_{{\widetilde{N}}_2}}\,
\frac{D_{21}(s)}{D_{11}(s)}\ +\ \dots\ ,\\[3mm]
  \label{D22exp}
\Delta_{22}(s) &=& \frac{Z_2(s)}{s - s_{{\widetilde{N}}_2}}\ +\
\frac{D_{21}(s)}{D_{22}(s)}\,\frac{Z_1(s)}{s - s_{{\widetilde{N}}_1}}\,
\frac{D_{12}(s)}{D_{22}(s)}\ +\ \dots\ ,\\[3mm]
  \label{D12exp}
\Delta_{12}(s) &=& -\,\frac{Z_1(s)}{s - s_{{\widetilde{N}}_1}}\,
\frac{D_{12}(s)}{D_{22}(s)} 
\ -\ \frac{D_{12} (s)}{D_{11} (s)}\,\frac{Z_2 (s)}{s - s_{{\widetilde{N}}_2}}\,
\ +\ \dots\ ,\\[3mm]
  \label{D21exp}
\Delta_{21}(s) &=& -\,\frac{Z_2(s)}{s - s_{{\widetilde{N}}_2}}\,
\frac{D_{21}(s)}{D_{11}(s)} 
\ -\ \frac{D_{21} (s)}{D_{22} (s)}\,\frac{Z_1 (s)}{s - s_{{\widetilde{N}}_1}}\,
\ +\ \dots\ ,
\end{eqnarray}
where the ellipses denote off-resonant terms which are non-singular at
$s=s_{{\widetilde{N}}_{1,2}}$.   Notice  that  we  have  retained  the
$s$-dependent analytic form for the residues in the above complex pole
expansion   by   virtue   of   the   Cauchy   theorem.    Substituting
(\ref{D11exp})--(\ref{D21exp})  into the  $s$-channel  amplitude $T_s$
in~(\ref{Ts}) and neglecting off-resonant terms yields
\begin{equation}
  \label{Tres}  
\widetilde{\cal   T}_s (s)\  =\ V^A_1   (s)\ \frac{Z_1
(s)}{s - s_{{\widetilde{N}}_1} }\ V^B_1(s)\ +\ V^A_2 (s)\ \frac{Z_2
  (s)}{s - s_{{\widetilde{N}}_2} }\ V^B_2(s)\, ,
\end{equation}
with 
\begin{equation}
 \label{VAB}
V_1^{A\, (B)}(s)\ =\ \Gamma_1^{A\, (B)}\ -\
\frac{D_{12}(s)}{D_{22}(s)}\ \Gamma_2^{A\, (B)} \,,\qquad
V_2^{A\, (B)}(s)\ =\ \Gamma_2^{A\, (B)}\ -\
\frac{D_{21}(s)}{D_{11}(s)}\ \Gamma_1^{A\, (B)} \; .
\end{equation}
Here, it  is important  to remark that  the expressions  $V_{1,2} (s)$
in~(\ref{VAB}) become  identical at $s  = m^2_{{\widetilde{N}}_{1,2}}$
to  the resummed  decay amplitudes  derived in~\cite{APRD1},  using an
LSZ-type reduction  formalism.  Instead, in the  present approach, the
corresponding resummed decay amplitude can be obtained by studying the
analytic structure of the residues of the complete resonant scattering
amplitude,  in which the  unstable heavy  sneutrinos are  described as
intermediate  states in  the  $s$-channel.  The  fact  that these  two
approaches lead to  identical results provides a firm  support for the
validity of the method developed in~\cite{APRD}.

The   RIS  squared   amplitude   pertinent  to   the  propagation   of
${\widetilde{N}}_1$ and ${\widetilde{N}}_2$ can now be identified as
\begin{eqnarray}
  \label{TRIS2}
|{\cal T}_{\rm RIS}(L\Phi \to L^C\Phi^\dagger)|^2 \!&=&\! \\[2mm]
&&\hspace{-2cm} \frac{|Z_1|^2 \pi\, \delta_+
  (s-m^2_{{\widetilde{N}}_1})}{m_{{\widetilde{N}}_1}
  \Gamma_{{\widetilde{N}}_1}}\,   
|V^A_1|^2\: |V^B_1|^2 \ +\ 
\frac{|Z_2|^2 \pi\, \delta_+
  (s-m^2_{{\widetilde{N}}_2})}{m_{{\widetilde{N}}_2}
\Gamma_{{\widetilde{N}}_2}}\,   
|V^A_2|^2\: |V^B_2|^2 \ ,\qquad\nonumber
\end{eqnarray}
with   $\delta_+    (s   -    m^2_{{\widetilde{N}}_{1,2}})   =  \delta
(s-m^2_{{\widetilde{N}}_{1,2}})\    \theta     (\sqrt{s})$.         In
(\ref{TRIS2}),   $Z_1$  ($Z_2$)   and   $V^{A,B}_1$ ($V^{A,B}_2$)  are
evaluated     at     $s     =   m^2_{{\widetilde{N}}_1}$     ($s     =
m^2_{{\widetilde{N}}_2}$).   Observe     that  up    to   higher-order
wave-function       renormalization  effects,    the    RIS    squared
amplitude~(\ref{TRIS2}) for two unstable  particles is very  analogous
to   the   corresponding   one~(\ref{TRIS})   derived for  one  single
resonance. Although we will not address this issue  in detail here, we
simply note that our subtraction approach of isolating the RIS part of
a squared  amplitude   can  be extended   to  more  than  two unstable
particles.    The  key observation to  be  made  here  is  that such a
generalization is  possible, since the $s$-channel dependent amplitude
in  the pole-dominance approximation can always  be expressed as a sum
of products  of resummed vertices  [cf.~(\ref{VAB})] and Breit--Wigner
propagators   with   single  complex  poles~[cf.~(\ref{Tsimple})   and
(\ref{Tres})].

One might worry that the  subtraction approach described above may not
be  applicable   for the   case    of our interest   with  overlapping
resonances,  i.e.\ for  $m_{N_2} -   m_{N_1} \sim \Gamma_{N_{1,2}}/2$.
However, we  should realize  that  the particles  associated  with the
complex  poles, e.g.\ $s_{N_{1,2}}$, of a  transition amplitude have a
completely  different    thermal history,    because  of    their many
decoherentional  collisions with the other   particles in the  thermal
bath.  On  the other  hand, the so-called  quantum  memory effects are
expected  to  play  a   relevant  role   only  when the  decay  widths
$\Gamma_{N_{1,2}}$ or the mass difference $m_{N_2} - m_{N_1}$ are much
smaller than the Hubble parameter $H$ governing  the expansion rate of
the early Universe at $T\approx m_{N_{1,2}}$.  In the former case, one
also   finds  that   $N_{1,2}$   are   weakly  thermalized~\cite{BBP}.
Otherwise, our subtraction approach  not only  takes into account  the
part of   the  squared amplitude associated  with  the  RIS,  but also
provides a consistent description  of the incoherent properties of the
heavy neutrinos~\footnote{For instance,  within the context of thermal
leptogenesis, the use of a  time-integrated CP-asymmetry formula, very
analogous     to   the one  applied     for   a coherently oscillating
$B^0\bar{B}^0$-system, leads  to   an erroneous incorporation  of  the
decoherentional properties of the thermal bath.}.

\subsection{Resummed Effective Yukawa Couplings and\\ 
                                                Leptonic Asymmetries}

Until now  in this section,  the heavy Majorana neutrinos  were mainly
treated as  scalar particles.   However, our approach  described above
for  subtracting the  RIS from  the squared  amplitude of  the process
$L\Phi  \to  L^C\Phi^\dagger$  carries  over  very  analogously  to  a
strongly-mixed fermionic system, including the spinorial nature of the
heavy Majorana neutrinos $N_1$ and~$N_2$. 

To see the above point, we first introduce an abbreviated form for the
one-loop corrected inverse $N_iN_j$-propagator matrix:
\begin{equation}
  \label{Dij}
\not\!\! D_{ij} (\not\! p)\ = \ \delta_{ij}\, (\not\! p - m_{N_i} )\: 
                                        +\: \Sigma_{ij} (\not\! p )\, ,
\end{equation}
where $\Sigma_{ij} (\not\!\!  p  )$ denote the self-energy transitions
$N_j (p) \to N_i (p)$, renormalized  in the OS  scheme, and $p$ is the
4-momentum of  $N_{1,2}$.   With  the  aid of   these newly-introduced
spinorial    functions~(\ref{Dij}), the resummed  $N_iN_j$-propagators
$S_{ij}(\not\!\! p)$ are given by (suppressing the argument $\not\! p$
everywhere)
\begin{eqnarray}
  \label{Sresum}
S_{11} &=& \Big( \not\!\!D_{11}\: -\: \not\!\!D_{12}\,
\not\!\!D^{-1}_{22}\, \not\!\!D_{21}\, \Big)^{-1}\,,\qquad
S_{22}\ \, = \ \, \Big( \not\!\!D_{22}\: -\: \not\!\!D_{21}\, 
\not\!\! D^{-1}_{11}\, \not\!\!D_{12}\, \Big)^{-1}\,,\nonumber\\
S_{12} &=& -\,S_{11}\: \not\!\!D_{12}\,\not\!\!D^{-1}_{22}\ = \ \
- \not\!\!D^{-1}_{11}\, \not\!\!D_{12}\, S_{22}\,,\nonumber\\
S_{21} &=& -\, S_{22}\: \not\!\!D_{21}\,\not\!\!D^{-1}_{11}\ = \ 
- \not\!\!D^{-1}_{22}\, \not\!\!D_{21}\, S_{11}\; ,\qquad
\end{eqnarray}
with $\not\!\!\!D^{-1}_{ij}(\not\!\!   p)   \  =  \  [\not\!\!\!D_{ij}
(\not\!\!p) ]^{-1}$.  These  expressions coincide with those presented
in~\cite{APRD}.  In~analogy to the  scalar case, we also introduce the
corresponding $Z$-factors:
\begin{equation}
  \label{Zf}
\not\!\!Z_i (\not\! p ) \ =\ \bigg(\,\frac{\partial}{\partial\!\not\!p}\, 
S^{-1}_{ii}(\not\! p )\,\bigg)^{-1}\; ,
\end{equation}
where the partial derivative  $\partial/\partial\!\not\!p$ may act  on
spinorial  expressions that depend on $\not\!p$  and  $p^2 {\bf 1}_4 =
(\not\! p)^2$. 

Since the  heavy-neutrino  self-energies $\Sigma_{ij} (\not\!  p)$ are
renormalized in   the OS scheme,  their  dispersive parts  satisfy the
renormalization conditions:
\begin{equation}
  \label{OSfermion}
{\rm Re}\, \Sigma_{ij} (\not\!  p)\; u_j (p)\ =\ 0\,,\qquad
\frac{1}{\not\! p - m_{N_i}}\ {\rm Re}\, \Sigma_{ii} (\not\!  p)\; u_i
(p)\ =\ 0\; .
\end{equation}
Again,  neglecting  contributions  from  unitarity  cuts,  it  can  be
shown~\cite{KP}   that    the   conditions~(\ref{OSfermion})   assure:
$\not\!\!Z^{\rm OS}_i (\not\!p)\; u_i(p)  = u_i (p)$ and $\bar{u}_i(p)
\not\!\!Z^{\rm OS}_i  (\not\!p) =  \bar{u}_i (p)$.  In  general, there
are deviations  from this last  equality, which result,  however, from
scheme-dependent, UV-finite terms of order $(h^\nu)^2$.

The complex pole positions of the resummed $N_iN_j$-propagators can be
determined by solving the equations:
\begin{equation}
  \label{poles}
D_i (s)\ =\ {\rm det}\, \Big[\, S^{-1}_{ii} (\not\! p) \, \Big]\ =\ 0\; ,
\end{equation}
where      the   determinant     is     taken     over   the spinorial
components~\footnote{To give  an   example,   we  note  that     ${\rm
det}\,(\not\!  p - m ) = (s - m^2)^2$, with  $s=p^2$. Observe that the
solutions  of ${\rm  det}\,(\not\!  p -  m )   =  0$ contain a  double
positive root at $\sqrt{s} = \sqrt{s}_{L,R}  = m$, reflecting the fact
that the  left- and right-handed components  of  a chiral field have
the same physical  pole as  a consequence  of  CPT invariance  of  the
theory.}. In fact, if ${\rm det}\, [\not\!\!  D_{ij} (\not\!  p)] \neq
0$ for  $i\neq  j$, it  is then sufficient   to solve  one of  the two
equations:  $D_1 (s) = 0$  or $D_2 (s)  = 0$, to  find the two complex
poles associated  with the unstable  heavy  neutrinos $N_1$ and $N_2$.
In  particular,   exactly as   in    the scalar  case, each   resummed
$N_iN_j$-propagator contains   two   complex  poles at   $\sqrt{s}   =
\sqrt{s}_{N_{1,2}}         =       m^{\rm       pole}_{N_{1,2}}\    -\
\frac{i}{2}\,\Gamma^{\rm pole}_{N_{1,2}}$.

By means  of Cauchy's  theorem, each resummed  $N_iN_j$-propagator can
now   be    expanded   about    the   complex   poles    $\sqrt{s}   =
\sqrt{s}_{N_{1,2}}$ as follows:
\begin{eqnarray}
  \label{S11exp}
S_{11}(\not\!p) &=& \frac{(\not\!p + m_{N_1})\, \not\!\!Z_1}{
s - s_{N_1}}\ +\
\frac{\not\!\!D^{-1}_{11}\, \not\!\!D_{12}\, (\not\!p + m_{N_2})\,
\not\!\!Z_2\, \not\!\!D_{21}\,\not\!\!D^{-1}_{11}}{s - s_{N_2}}\ 
+\ \dots\ ,\nonumber\\[3mm]
  \label{S22exp}
S_{22} (\not\!p) &=& \frac{ (\not\!p + m_{N_2})\, \not\!\!Z_2}
{s - s_{N_2}}\ +\
\frac{\not\!\!D^{-1}_{22}\,\not\!\! D_{21}\, (\not\!p + m_{N_1} )\, 
\not\!\!Z_1\,\not\!\!D_{12}\,\not\!\!D^{-1}_{22}}{s - s_{N_1}}\ 
+\ \dots\ ,\nonumber\\[3mm]
  \label{S12exp}
S_{12}(\not\!p) &=& -\,\frac{(\not\!p + m_{N_1})\,\not\!\!Z_1\, 
\not\!\!D_{12}\,\not\!\!D^{-1}_{22} }{s - s_{N_1}}
\ -\ \frac{\not\!\!D^{-1}_{11}\, \not\!\!D_{12}\, 
(\not\!p + m_{N_2})\, \not\!\!Z_2}{s - s_{N_2}}\,
\ +\ \dots\ ,\nonumber\\[3mm]
  \label{S21exp}
S_{21}(\not\!p) &=& -\,\frac{(\not\!p + m_{N_2})\,\not\!\!Z_2\, 
\not\!\!D_{21}\,\not\!\!D^{-1}_{11}}{s - s_{N_2}}
\ -\ \frac{\not\!\!D^{-1}_{22}\, \not\!\!D_{21}\, 
(\not\!p + m_{N_1})\, \not\!\!Z_1 }{s - s_{N_1}}\,
\ +\ \dots\ ,
\end{eqnarray}
where the explicit dependence of $\not\!\!Z_i$ and $\not\!\!D_{ij}$ on
$\not\!\!p$  is not shown.    ~~As before, the  ellipses indicate that
non-singular terms  at  $s=s_{N_{1,2}}$ have  been omitted as  well as
higher-order scheme-dependent terms beyond the OS renormalization.

After inserting  the pole-expanded expressions~(\ref{S11exp}) into the
transition  amplitude for the process  $L\Phi \to L^C\Phi^\dagger$, we
find very analogously to the scalar case:
\begin{equation}
  \label{Tresf}  
\widetilde{\cal   T}_s (s)\  =\ \overline{\not\!V}^A_1\ 
\frac{(\not\!p + m_{N_1})\,\not\!\!Z_1}{s - s_{N_1} }\ \not\!V^B_1\ +\ 
\overline{\not\!V}^A_2 \ \frac{(\not\!p + m_{N_2})\, 
\not\!\!Z_2}{s - s_{N_2}}\  
\not\! V^B_2\, ,
\end{equation}
with 
\begin{equation}
 \label{VABf}
\not\!V_1^{A\, (B)}\ =\ \Gamma_1^{A\, (B)}\ -\
\not\!\!D_{12} \not\!\!D^{-1}_{22}\, \Gamma_2^{A\, (B)} \,,\qquad
\not\!V_2^{A\, (B)}\ =\ \Gamma_2^{A\, (B)}\ -\
\not\!\!D_{21} \not\!\!D^{-1}_{11}\, \Gamma_1^{A\, (B)} \; ,
\end{equation}
and     $\overline{\not\!       V}^{A\,    (B)}_{1,2}      =    \not\!
V^{A\,(B)\,\dagger}_{1,2}    \gamma_0$.  The incoherent subtraction of
the   RIS    from   the  squared     amplitude  $|{\cal    T}(L\Phi\to
L^C\Phi^\dagger)  |^2$  can   be consistently    performed, after  the
de-correlation effect of the  heavy-neutrino  spins discussed in   the
previous   subsection  has  been  taken   into  account  by   means of
(\ref{Fierz}).  Then,  in the  pole-dominance approximation, we obtain
for $|{\cal   T}_{\rm RIS}(L\Phi \to  L^C\Phi^\dagger)|^2$  a  formula
analogous to~(\ref{TRIS2}), where  $|V_{1,2}^{A\, (B)}|^2$ is replaced
by  $${\rm  Tr}\, [\overline{\not\!    V}^{A\, (B)}_{1,2}   (\not\!p +
m_{N_{1,2}}) \not\!V_{1,2}^{A\,  (B)}]\; .$$ Finally, we should stress
again that these results are  in complete agreement with those derived
by the LSZ-type resummation approach in~\cite{APRD}.

In either resummation approach,  i.e.~the LSZ-type approach or the one
followed here, the   resummed  effective  amplitudes  for the   decays
$N_{1,2}(p)\to L \Phi$, denoted as  ${\cal T}_{N_{1,2}}$, are uniquely
determined by
\begin{eqnarray}
  \label{TN}
{\cal T}_{N_1} \!\!&=&\!\! \bar{u}_l P_R\, \Big\{ 
h^\nu_{l1}+i{\cal V}^{\rm abs}_{l1} (\not\! p) -
i\Big[h^\nu_{l2}+ i{\cal V}^{\rm abs}_{l2} (\not\! p) \Big]\,
\Big[\not\! p - m_{N_2} + i\Sigma_{22}^{\rm abs}(\not\! p)\Big]^{-1} 
\Sigma_{21}^{\rm abs}(\not\! p)\Big\} u_{N_1}(p)\; ,\nonumber\\
{\cal T}_{N_2} \!\!&=&\!\! \bar{u}_l P_R\, \Big\{ 
h^\nu_{l2}+i{\cal V}^{\rm abs}_{l2} (\not\! p) -
i\Big[h^\nu_{l1}+ i{\cal V}^{\rm abs}_{l1} (\not\! p) \Big]\,
\Big[\not\! p - m_{N_1} + i\Sigma_{11}^{\rm abs}(\not\! p)\Big]^{-1} 
\Sigma_{12}^{\rm abs}(\not\! p)\Big\} u_{N_2}(p)\; ,\nonumber\\
\end{eqnarray}
where one-loop contributions   from the vertices  $N_{1,2}\to  L \Phi$
have also been included.   ~~In~writing~(\ref{TN}), we have implicitly
assumed that all  Yukawa couplings and masses  are renormalized in the
OS scheme.  Then, up to  higher-order scheme-dependent terms, only the
absorptive   parts ${\cal   V}^{\rm    abs}_{li}   (\not\!  p)$    and
$\Sigma_{ij}^{\rm abs}(\not\!    p)$  of  the  one-loop   vertices and
self-energies become relevant. These are given by~\cite{APRD}
\begin{eqnarray}
  \label{Sabs}
\Sigma^{\rm abs}_{ij} (\not\! p) \!&=&\! A_{ij} \not\! p\, P_L\ +\ 
A^*_{ij} \not\! p\, P_R\ =\ 
\sum\limits_{l'=1}^3\ \bigg(\,
\frac{h^\nu_{l'i}\,h^{\nu *}_{l'j}}{16\pi} \not\! p\, P_L\ +\ 
\frac{h^{\nu *}_{l'i}\,h^\nu_{l'j}}{16\pi} \not\! p\, P_R\,\bigg)\, ,\\
  \label{Vabs}
{\cal V}^{\rm abs}_{li}(\not\! p) \!&=&\!
\frac{B_{li}}{\sqrt{p^2}} \not\!p\,P_L
\ =\  -\,
\sum\limits_{l'=1}^3\ \sum\limits_{\stackrel{j=1,2}{{}^{(j\neq i)}}}\ 
\frac{h^{\nu *}_{l'i}\,h^\nu_{l'j}\,h^\nu_{lj}}{16\pi\sqrt{p^2}} 
\not\!p\, P_L\, f\bigg(\frac{m^2_{Nj}}{p^2}\bigg)\, ,
\end{eqnarray}
where $A_{ji} =  A^*_{ij}$  and  $f(x)=\sqrt{x}[1-(1+x)\ln(1+1/x)]$ is
the Fukugita--Yanagida one-loop function~\cite{FY}. 

Substituting~(\ref{Sabs})      and~(\ref{Vabs}) into~(\ref{TN})    and
neglecting terms which are formally   of order $(h^\nu)^4$ and  higher
yields
\begin{equation}
  \label{Thres}
{\cal T} (N_i \to L\Phi) \ =\ 
 (\bar{h}^\nu_+)_{li}\ \bar{u}_l\, P_R\, u_{N_i}\; ,
\end{equation}
with
\begin{eqnarray}
  \label{hres}
(\bar{h}^\nu_+)_{l1} \!&=&\! 
h^\nu_{l1}\, +\, iB_{l1}\: -\: \frac{i h^\nu_{l2}\, m_{N_1}\, 
( m_{N_1}\, A_{12}\: +\: m_{N_2}\, A_{21} )}{m^2_{N_1}\, -\, 
m^2_{N_2}\, +\, 2i\,A_{22}\,m^2_{N_1} }\ ,\nonumber\\ 
(\bar{h}^\nu_+)_{l2} \!&=&\! 
h^\nu_{l2}\, +\, iB_{l2}\: -\: \frac{i h^\nu_{l1}\, m_{N_2}\, 
( m_{N_2}\, A_{21}\: +\: m_{N_1}\, A_{12} )}{m^2_{N_2}\, -\, 
m^2_{N_1}\, +\, 2i\,A_{11}\,m^2_{N_2} }\ ,
\end{eqnarray}
where the parameters $A_{ij}$ and $B_{li}$ are defined in~(\ref{Sabs})
and (\ref{Vabs}).  The  CP-conjugate decay amplitudes ${\cal T}(N_i\to
L^C\Phi^\dagger )$ can  easily be recovered from~(\ref{Thres}) by just
taking the complex-conjugate Yukawa couplings in~(\ref{hres}), i.e.
\begin{eqnarray}
  \label{hresC}
(\bar{h}^\nu_-)_{l1} \!&=&\! 
h^{\nu *}_{l1}\, +\, iB^*_{l1}\: -\: \frac{i h^{\nu *}_{l2}\, m_{N_1}\, 
( m_{N_1}\, A^*_{12}\: +\: m_{N_2}\, A^*_{21} )}{m^2_{N_1}\, -\, 
m^2_{N_2}\, +\, 2i\,A_{22}\,m^2_{N_1} }\ ,\nonumber\\ 
(\bar{h}^\nu_-)_{l2} \!&=&\! 
h^{\nu *}_{l2}\, +\, iB^*_{l2}\: -\: \frac{i h^{\nu *}_{l1}\, m_{N_2}\, 
( m_{N_2}\, A^*_{21}\: +\: m_{N_1}\, A^*_{12} )}{m^2_{N_2}\, -\, 
m^2_{N_1}\, +\, 2i\,A_{11}\,m^2_{N_2} }\ .
\end{eqnarray}
Notice that as a  consequence of CP  violation, it  is $|(\bar{h}^{\nu
}_+)^*_{li}|   \neq  |(\bar{h}^\nu_-)_{li}|$.   In particular,   it is
important to     remark    that   the effective     Yukawa   couplings
$\bar{h}^\nu_\pm$ defined  in~(\ref{hres})  and~(\ref{hresC})  contain
the dominant part of the one-loop radiative corrections as well as the
enhanced heavy-neutrino self-energy  effects  in the  kinematic region
$(m_{N_2} - m_{N_1})  \ll (m_{N_1} +  m_{N_2})$.  As we will see below
and    in Appendix~\ref{app:CT}, the    leptonic  asymmetries  and the
radiatively-corrected collision terms  can  be expressed in  a compact
manner   in  terms   of the    resummed  effective Yukawa    couplings
$\bar{h}^\nu_\pm$.

\begin{figure}[t]
\begin{center}
\begin{picture}(300,100)(0,0)
\SetWidth{0.8}

\Vertex(50,50){2}
\Line(50,50)(90,50)\Text(70,62)[]{$N_i$}
\Line(130,50)(170,50)\Text(150,62)[]{$N_j$}
\GCirc(110,50){20}{0.9}\Text(110,50)[]{{\boldmath $\varepsilon$}}
\GCirc(180,50){10}{0.9}\Text(180,50)[]{{\boldmath $\varepsilon'$}}
\DashArrowLine(187,55)(220,80){5}\Text(225,80)[l]{$\Phi$}
\ArrowLine(187,45)(220,20)\Text(225,20)[l]{$L$}

\end{picture}
\end{center}
\caption{$\varepsilon$- and  $\varepsilon'$-types of CP violation
in the decays of heavy Majorana neutrinos.}\label{fig3}
\end{figure}
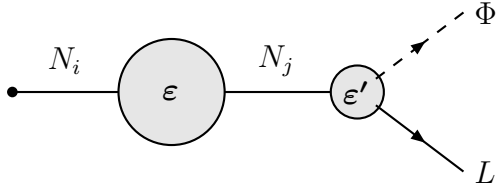

Let  us  now  turn our attention  to   the discussion of  the leptonic
asymmetries~$\delta_{N_{1,2}}$.       The   CP-violating    quantities
$\delta_{N_{1,2}}$ are defined and easily calculated as
\begin{equation}
  \label{deltaNi}
\delta_{N_i}\ =\ \frac{\Gamma (N_i\to L\Phi )\, -\, 
\Gamma (N_i\to L^C \Phi^\dagger)}{\Gamma (N_i\to L\Phi )\, +\, 
\Gamma (N_i\to L^C \Phi^\dagger)}\ =\ 
\frac{(\bar{h}^{\nu\,\dagger}_+ \bar{h}^\nu_+)_{ii} \: - \: 
(\bar{h}^{\nu\,\dagger}_- \bar{h}^\nu_-)_{ii}}{
(\bar{h}^{\nu\,\dagger}_+ \bar{h}^\nu_+)_{ii} \: + \: 
(\bar{h}^{\nu\,\dagger}_- \bar{h}^\nu_-)_{ii}}\ .
\end{equation}
In~(\ref{deltaNi}),  a  matrix  notation  for the  resummed  effective
Yukawa couplings  $(\bar{h}^\nu_\pm)_{li}$ should be  understood, with
$l=1,2,3$ and $i = 1,2$.   As is illustrated in Fig.~\ref{fig3}, it is
now interesting  to discuss  the two different  types of  CP violation
contributing  to  $\delta_{N_i}$,  the so-called  $\varepsilon'$-  and
$\varepsilon$- types  of CP violation.  If we  neglect all self-energy
contributions    to~$\delta_{N_i}$    and   ${\cal    O}[(h^{\nu})^3]$
CP-conserving  terms,   we  then  find   the  known  result   for  the
$\varepsilon'$-type CP violation~\cite{FY}:
\begin{equation}
  \label{eps'N}
\delta_{N_i}\ \approx\ \varepsilon'_{N_i}\ =\ 
\frac{ {\rm Im}\, (h^{\nu\dagger}\,h^\nu)^2_{ij}}{
8\pi (h^{\nu\dagger}\,h^\nu)_{ii}}\ 
f\bigg(\frac{m^2_{N_j}}{m^2_{N_i}}\bigg)\, ,
\end{equation}
with $i  \neq j$.   Instead, if all  one-loop vertex  corrections have
been neglected,  we obtain a simple formula  for the $\varepsilon$-type CP
violation~\cite{APRD,APreview}:
\begin{equation}
  \label{epsN}
\delta_{N_i}\ \approx\ 
\varepsilon_{N_i}\ =\ \frac{{\rm Im}\, (h^{\nu\dagger}\,h^\nu)^2_{ij}}{
(h^{\nu\dagger}\,h^\nu)_{ii}\,(h^{\nu\dagger}\,h^\nu)_{jj} }\ 
\frac{ (m^2_{N_i} - m^2_{N_j})\, m_{N_i}\, \Gamma^{(0)}_{N_j} }{
(m^2_{N_i} - m^2_{N_j})^2\, +\, m^2_{N_i}\Gamma^{(0)\,2}_{N_j}}\ ,
\end{equation}
where  $i,j=1,2$  $(i  \neq   j)$  and  the  tree-level  decay  widths
$\Gamma^{(0)}_{N_i}$ of  the heavy Majorana neutrinos  $N_i$ are given
in~(\ref{GNi}).  In  finite-order perturbation theory,  the absorptive
term  $m^2_{N_i}  \Gamma^{(0)\,2}_{N_j}$   that  occurs  in  the  last
denominator on the RHS of~(\ref{epsN}) is absent, thereby leading to a
singular  behaviour  for  $\varepsilon_{N_i}$ in  the  mass-degenerate
limit   $m_{N_i}\to  m_{N_j}$.   However,   the  appearance   of  this
regulating  absorptive term  due  to  the finite  width  of the  heavy
Majorana neutrinos should be  expected on physical grounds and emerges
naturally  within our  resummation  approach.  Finally,  it should  be
noted that  (\ref{epsN}) is valid for  a mixing system  with two heavy
Majorana neutrinos  only.  The generalization  of~(\ref{epsN}) for the
case of  a three-heavy-Majorana-neutrino  mixing is more  involved and
hence has been relegated to Appendix~\ref{app:mixing}.

A  non-zero leptonic  asymmetry can be   created,  if and only if  the
following CP-odd quantity does not vanish~\cite{APRD}:
\begin{equation}
  \label{CPodd} 
\Delta_{\rm CP}\ \equiv\ {\rm Im}\, {\rm Tr}\, \bigg[\,
(h^{\nu_R})^\dagger\, h^{\nu_R}\, M^\dagger_S\, M_S\, M^\dagger_S\,
(h^{\nu_R})^T\, (h^{\nu_R})^*\, M_S\,\bigg] \ \neq\ 0\; ,
\end{equation}
where $h^{\nu_R}$  and $M_S$ are  the Yukawa-coupling and  the singlet
neutrino-mass     matrices,    defined     in    the     weak    basis
through~(\ref{LYint}) and (\ref{Majmass}), respectively.  An important
property  of the  CP-odd  quantity  $\Delta_{\rm CP}$  is  that it  is
invariant      under       the      so-called      generalized      CP
transformations~\cite{BBG,Branco}. In the physical basis, $\Delta_{\rm
CP}$ is found to be
\begin{eqnarray}
  \label{2genCP}
\Delta^{\rm 2G}_{\rm CP} \!& = &\! 
m_{N_1}\, m_{N_2}\, (m^2_{N_2} - m^2_{N_1})\,
{\rm Im}\, \Big[ (h^{\nu\dagger}\,h^\nu)^2_{12}\Big]\;, \\[3mm]
  \label{3genCP}
\Delta^{\rm 3G}_{\rm CP} \!& = &\! 
m_{N_1}\, m_{N_2}\, (m^2_{N_2} - m^2_{N_1})\,
{\rm Im}\, \Big[ (h^{\nu\dagger}\,h^\nu)^2_{12}\Big]\: +\: 
m_{N_2}\, m_{N_3}\, (m^2_{N_3} - m^2_{N_2})\,
{\rm Im}\, \Big[ (h^{\nu\dagger}\,h^\nu)^2_{23}\Big]\nonumber\\
\!&&\! +\, m_{N_1}\, m_{N_3}\, (m^2_{N_3} - m^2_{N_1})\,
{\rm Im}\, \Big[ (h^{\nu\dagger}\,h^\nu)^2_{13}\Big]\; ,
\end{eqnarray}
for a two- and  a three-generation heavy-neutrino model, respectively.
In  general,  the total  number  ${\cal  N}_{CP}$  of all  non-trivial
CP-violating phases in  a model with $n_L$ weak  isodoublets and $n_R$
neutral singlets is ${\cal N}_{CP} = n_L(n_R-1)$~\cite{KPS}.  However,
after summing over all lepton  flavours that occur in the final states
of heavy Majorana-neutrino decays, only one CP-violating phase becomes
relevant for leptogenesis which can be equivalently represented by the
rephasing-invariant quantity $\Delta_{\rm CP}$ in~(\ref{CPodd}).

Technically, we may understand this last point as follows.  As we will
explicitly see in  Section~4.2,   the  net  source $\delta n_L$    for
generating a non-zero value for the number density $n_L$ of the lepton
number $L$ has the analytical structure:
\begin{equation}
  \label{dnL}
\delta n_L\ =\ \sum\limits_{i=1,2,3}\, \delta_{N_i}\,\Gamma_{N_i}\
                                                          g(m_{N_i})\; ,
\end{equation}
where  $\Gamma_{N_{1,2,3}}$  are the radiatively-corrected total decay
widths of $N_i$ and $g (m_{N_i})$ is an analytic function of $m_{N_i}$
that  contains Boltzmann-type factors, and whose   precise form is not
important  for the present   discussion.    Since $\delta n_L$  is   a
physical CP-violating     quantity in leptogenesis,    it  should   be
proportional  to       $\Delta_{\rm  CP}$.   Indeed,      it  can   be
straightforwardly  checked  that in ordinary finite-order perturbation
theory, both $\varepsilon$- and $\varepsilon'$-  types of CP violation
are  proportional to $\Delta_{\rm CP}$, only  after the sum over $N_i$
is   considered~\footnote{This statement   is    only  valid   for   a
three-generation  heavy-neutrino  model.  For a  model  with two heavy
Majorana  neutrinos,  the   $\varepsilon$-type CP-violating  terms  in
$\delta_{N_i}$ are individually  proportional to $\Delta^{\rm 2G}_{\rm
CP}$.}.  Most remarkably, it has been shown for a model with two heavy
Majorana neutrinos~\cite{APRD}  that even the  resummed expressions of
$\delta_{N_i}$ given in~(\ref{deltaNi})  lead to a $\delta n_L \propto
\Delta^{\rm  2G}_{\rm CP}$;  the three-heavy-Majorana-neutrino case is
discussed in  Appendix~\ref{app:mixing}.    This property  shows   the
consistency of our resummation approach with respect to generalized CP
transformations.

\setcounter{equation}{0}
\section{Boltzmann Equations for Resonant Leptogenesis}\label{sec:BEs}

Before we solve numerically  the relevant BEs,  it is useful to give a
qualitative discussion  of the  out-of-equilibrium constraints  on the
parameters of the theory. Specifically, within the context of resonant
leptogenesis,  we  demonstrate  how  moderate   departures  from   the
out-of-equilibrium condition on the  decay rates of the heavy Majorana
neutrinos are sufficient  to significantly lower  the singlet scale to
the TeV    range,  without being   in  conflict   with  neutrino data.
Subsequently,  we set up the  relevant network of BEs, where important
contributions  to scatterings from enhanced heavy-neutrino self-energy
graphs and from gauge-mediated   interactions are  included.   Solving
numerically the BEs  for  specific scenarios compatible with   neutrino
data, we show that the leptogenesis scale can be lowered up to the TeV
scale or even lower close to the critical temperature $T_c$, namely up
to the scale at which the $(B+L)$-violating sphaleron interactions are
still in thermal equilibrium.

\subsection{Out-of-Equilibrium Constraints}

The  out-of-thermal    equilibrium condition   on  the  heavy Majorana
neutrino decays   places severe  limits on  the  Yukawa  couplings  of
neutrino  models~\cite{KT}.  To obtain a qualitative  understanding of
those limits, let us first introduce the parameters
\begin{equation}
\label{Ki}
K_i\ =\ \frac{\Gamma^{(0)}_{N_i}}{H(T=m_{N_i})}\ \sim\
\frac{\Gamma_{N_i}}{H(T=m_{N_i})}
\end{equation}
where  $\Gamma^{(0)}_{N_i}$ and $\Gamma_{N_i}$  are the tree-level and
resummed  total decay  widths    of $N_i$,  respectively,  and  $H(T)$
in~(\ref{Ki}) is the Hubble parameter
\begin{equation}
\label{Hubble}
H(T)\ =\ 1.66\, g_*^{1/2}\, \frac{T^2}{M_{\rm Planck}}\ .
\end{equation}
In~(\ref{Hubble}), $M_{\rm  Planck} =   1.2\times 10^{16}$~TeV is  the
Planck mass and $g_*\approx 107$ is the number of relativistic degrees
of freedom of the SM.   Obviously,  the parameters $K_i$ quantify  the
deviation of  the heavy-neutrino decay rates $\Gamma^{(0)}_{N_i}$ from
the  expansion rate of the Universe.    These parameters should not be
much  larger than a certain maximum  value  $K_i^{\rm max}$, such that
leptogenesis can be  successfully realized.  Even though most analyses
conservatively assume $K_i^{\rm  max} \sim 1$,  much larger  values of
$K_i^{\rm max}$ even larger  than  1000 can  still be  tolerated.   In
Section~\ref{sec:BEs}.3, we  provide    a    firm support    of   this
observation, after numerically solving  the corresponding BEs.

Let   us   now  consider   the   out-of-equilibrium  constraint   $K_i
\stackrel{<}{{}_\sim} K_i^{\rm  max}$. This is  easily translated into
the upper bound
\begin{equation}
\label{hli_bound}
(h^{\nu\dagger} h^\nu)_{ii}\ \stackrel{<}{{}_\sim}\ 
3.5\, K^{\rm max}_i \times 10^{-14}\ 
\bigg(\frac{m_{N_i}}{1\ \mbox{TeV}}\bigg)\, .
\end{equation}
The very same upper bound can  be expressed in terms of new parameters
$\widetilde{m}_i$, called effective neutrino masses in~\cite{BBP}, i.e.
\begin{equation}
  \label{meff_bound}
\widetilde{m}_i\ \equiv\ \frac{v^2\, (h^{\nu\dagger} h^\nu)_{ii}}{2\, m_{N_i}}\
\stackrel{<}{{}_\sim}\ 10^{-3}\, K^{\rm max}_i~{\rm eV}\, .
\end{equation}

At temperatures $T$ above  the electroweak phase transition, i.e.\ for
$T\stackrel{>}{{}_\sim}   T_c   \approx  200$~GeV,   $(B+L)$-violating
interactions mediated by sphalerons  are in thermal equilibrium.  This
in-equilibrium condition  gives rise to the  following relations among
the number-to-entropy ratios of densities~\cite{BS,HT}:
\begin{equation}
  \label{conversion}
Y_B (T > T_c)\ =\ \frac{28}{79}\ Y_{B-L} (T > T_c)\ =\ 
-\frac{28}{51}\, Y_L (T > T_c)\,,
\end{equation}
where  we have defined
\begin{equation}
  \label{Yx}
Y_X\ =\  \frac{n_X}{s}\ ,
\end{equation}
with   $X=B,L,(B-L)$,  and   $s$  is   the  entropy   density.   Thus,
approximately one-half  of the lepton-to-entropy density  ratio $Y_L =
n_L/s$ gets  converted into a  baryon-to-entropy density ratio  $Y_B =
n_B/s$   through    the   equilibrated   $(B+L)$-violating   sphaleron
interactions.

In order to relate the $Y_B$  generated at a temperature $T=T_* > T_c$
to $\eta_B$ measured  much later at the recombination  epoch $T_0$, we
may  conveniently assume  that there  is  no source  or mechanism  for
significant entropy  release while the  Universe is cooling  down from
$T_*$  to   $T_0$.   Under   this  plausible  assumption   of  entropy
conservation, one  can then  establish the relation  $Y_B (T_*)  = Y_B
(T_0)$. Employing now the fact that  $s (T) = g_s (T) n_\gamma (T)$ is
the entropy  density of a plasma  with a number  $g_s$ of relativistic
degrees of freedom at temperature $T$, we arrive at the relation:
\begin{equation}
  \label{etaBth}
\eta_B\ =\  \frac{g_s (T_0)}{g_s (T_*)}\ \frac{n_B(T_*)}{n_\gamma(T_*)}\ .
\end{equation}
For our  numerical analysis, we use~\cite{KT,BBP}: $g_s  (T_0) = 3.91$
and $g_s  (T_*) = 107$,  i.e.\ $g_s (T_0)/g_s (T_*)\approx  1/27$.  If
$K_i < K_i^{\rm max}$, an order-of-magnitude estimate for $\eta_B$ may
be obtained by
\begin{equation}
  \label{etaBapprox} 
\eta_B\  \sim\  -\!   \sum\limits_{i=1,2,3}\  \frac{\delta_{N_i}}{200\,
K_i}\     \approx\    -\!      \sum\limits_{i=1,2,3}\    \frac{1}{200}\
\bigg(\frac{10^{-3}~{\rm eV}}{\widetilde{m}_i}\bigg)\ \delta_{N_i}\ .
\end{equation}
In the  above formula, the lepton-to-baryon  conversion factor through
sphalerons,  given in~(\ref{conversion}),  has also  been implemented.
It is  not difficult to see from~(\ref{etaBapprox})  that $\eta_B$ can
be   around   the   observed    value   of   $6\times   10^{-10}$   if
$|\delta_{N_i}|/K_i$ is of  order $10^{-8}$. Evidently, CP asymmetries
of order unity  allow for very large values of  $K_i$.  In fact, large
values  of $K_i  \gg  1$ lead  to  a thermally  dense  plasma, so  the
required  conditions of  kinetic  equilibrium and  decoherence of  the
heavy  Majorana  neutrinos  in  the  BEs  are  comfortably  satisfied.
Furthermore, since most of the heavy neutrinos have already decayed at
$T\approx m_{N_i}\approx  m_N$, there are no  dilution factors through
entropy production at lower temperatures $T\ll m_N$~\cite{MAL}.
 
It is  now straightforward  to examine whether  the out-of-equilibrium
constraints on  the Yukawa couplings~(\ref{hli_bound}),  together with
the  estimate~(\ref{etaBapprox}) for  a successful  generation  of the
BAU, still allow a  light-neutrino sector that can adequately describe
the  solar and  atmospheric neutrino  data.  In  the framework  of the
generic   models    discussed   in   Section~\ref{sec:models},   their
hierarchical   light-neutrino  mass   spectrum  requires   that  ${\rm
Tr}\,({\bf  m}^\nu) \approx  0.05$~eV,  where ${\bf  m}^\nu$ is  given
by~(\ref{mnuappr}). This implies that
\begin{equation}
  \label{hbound}
\bigg| \sum\limits_{i=1,2,3}\, h_{i1}\,h_{i2}\, \bigg|\ 
\approx\ 10^{-12}\, \bigg(\,\frac{m_N}{{\rm TeV}}\,\bigg)\; .
\end{equation}
Comparing   (\ref{hli_bound})  and  (\ref{hbound})   and  assuming  no
accidental cancellations in the different sums of Yukawa couplings, we
find  that values of $K_i$  larger  than $\sim  30$ are sufficient  to
successfully   describe  the  solar  and  atmospheric  neutrino  data,
provided $K_i^{\rm  max}\stackrel{>}{{}_\sim}  100$.  Most remarkably,
this  result is  almost  independent of the  leptogenesis scale $m_N$,
which can be as   low as 1~TeV.    A rigorous demonstration   of this
observation will be given in Section~\ref{sec:BEs}.3.

In  our discussion  above,  we have  assumed  that finite  temperature
effects  will  not  affect  significantly  the  main  results  of  our
analysis. In particular,  one may have to worry  whether the condition
for  resonant CP  violation stated  in~(\ref{CPres})  will drastically
modify  under  the  influence  of  thermal  effects.   Indeed,  finite
temperature  effects on  the  $T=0$  masses of  the  SM particles  are
significant.   Gauge and  top-quark Yukawa  interactions give  rise to
appreciable   thermal   masses  for   the   leptons   and  the   Higgs
fields~\cite{HAW}, i.e.
\begin{eqnarray}
  \label{thermal}
\frac{m^2_L(T)}{T^2} &=& \frac{1}{32}\, (3g^2\, +\, g^{\prime 2})\, ,
                                                             \nonumber\\
\frac{M^2_\Phi (T)}{T^2} &=& 2\,d\, \bigg(\,1\: -\: 
                                             \frac{T^2_c}{T^2}\,\bigg)\;,
\end{eqnarray}
where $g$ and  $g'$ are the SU(2)$_L$ and  U(1)$_Y$ gauge couplings at
the  running  scale  $T$,  and  $d  =  (8M^2_W  +  M^2_Z  +  2m^2_t  +
M^2_H)/(8v^2)$.  At  temperatures $T\stackrel{<}{{}_\sim} m_{N}$ where
leptogenesis becomes  more operative, thermal mass  effects on leptons
are roughly  one order  of magnitude smaller  than the  heavy Majorana
neutrino mass. However, thermal contributions to the mass of the Higgs
field are  more important, and highly depend~\cite{CKO}  on the actual
value of the $T=0$ Higgs-boson mass $M_H$.  If the range of Higgs-mass
values $115~{\rm  GeV} \stackrel{<}{{}_\sim} M_H \stackrel{<}{{}_\sim}
190$~GeV is  considered, which is  deduced from direct  Higgs searches
and  electroweak precision  data, one  then gets  the upper  and lower
limits: $0.5  \stackrel{<}{{}_\sim} M_\Phi (T)/T \stackrel{<}{{}_\sim}
0.7$. As a  result, the effective decay widths  of the heavy neutrinos
$\Gamma_{N_i}(T=m_{N_i})$  will reduce at  most by  a factor  $\sim 2$
with respect to $\Gamma_{N_i}(T=0)$ due to phase-space corrections.

Instead, thermal  effects on  heavy Majorana-neutrino masses  are very
suppressed,  as they  are  proportional to  the heavy-neutrino  Yukawa
couplings  $h^\nu_{ij}$.  Adapting  the results  of~\cite{HAW}  to our
model, the size of these thermal effects may be computed by
\begin{equation}
  \label{mN(T)}
\frac{\widehat{M}^2_S (T)\, -\, \widehat{M}^2_S (0)}{T^2}\ =\ \frac{1}{16}\, 
(h^{\nu\dagger} h^\nu )_{ij}\, ,
\end{equation}
where $\widehat{M}_S (0)$ is the physical diagonal heavy-neutrino mass
matrix defined in~(\ref{Utrans}) in  the symmetric phase of the theory
at $T=0$. At finite  temperatures, $\widehat{M}^2_S (T)$ is in general
not diagonal  and therefore needs  a $T$-dependent re-diagonalization.
However,  from~(\ref{mN(T)})  it  can  be estimated  that  for  nearly
degenerate heavy  Majorana neutrinos  with $m_{N_i} \approx  m_N$, the
thermally induced mass splitting is
\begin{equation}
  \label{thsplit}
m_{N_i}(T)\ -\ m_{N_j} (T)\ \stackrel{<}{{}_\sim}\ \frac{1}{16}\
{\rm Re}\,[(h^{\nu\dagger} h^\nu )_{ij}]\ \frac{T^2}{m_N}\; .
\end{equation}
This thermally induced   mass splitting is   comparable to  the  decay
widths $\Gamma_{N_i}$ of the  heavy Majorana neutrinos at temperatures
$T\stackrel{<}{{}_\sim} m_N$, at which a net $L$ and $B$ number can in
principle  be created.  As a  consequence, the conditions for resonant
CP violation in~(\ref{CPres}) are not spoiled  by thermal effects in a
relevant way.

\subsection{Boltzmann Equations}

Our derivation of  the BEs relies  on  a number of approximations  and
valid simplifications.   In particular, we  neglect thermal effects on
all  collision terms  which become  less  significant for temperatures
$T\stackrel{<}{{}_\sim}  m_{N_1}$  relevant    to   leptogenesis.   In
addition,  we  adopt  the  Maxwell--Boltzmann   statistics, which   is
expected  to introduce  errors no   larger than  20\%.   Nevertheless,
several crucial improvements,  which were  neglected in the   existing
literature, have now been implemented in the BEs.   To be specific, we
include  the gauge-mediated collision   terms that describe  processes
such as $N_i  L_j\!  \leftrightarrow\!  \Phi^\dagger V_\mu$  and their
crossing-symmetric reactions, where  $V_\mu$ collectively denotes  the
SU(2)$_L$ and  U(1)$_Y$  gauge bosons  $W^a_\mu$ (with  $a=1,2,3$) and
$B_\mu$ in the unbroken  symmetric phase  of the  SM.  Since our  main
interest is  resonant     leptogenesis, we  include     heavy-neutrino
self-energy  enhanced contributions  to  scatterings according to  the
resummation approach analyzed in Section~\ref{sec:RIS}.

Before  writing down  the relevant  set of  the BEs,  it is  useful to
establish notation  and define a number of  auxiliary quantities.  For
this purpose,  let us  start by reviewing  a few basic  concepts.  The
number  density $n_a$  of particle  species $a$,  with  $g_a$ internal
degrees of freedom, is given by~\cite{KW}
\begin{eqnarray}
  \label{na}
n_a (T) &=& g_a\, \int \frac{d^3{\bf p}}{(2\pi)^3}\ 
\exp\Big[ - \Big(\sqrt{{\bf p}^2 + m^2_a} -  \mu_a (T)\Big)/T\,\Big]\nonumber\\
&=&  \frac{g_a\, m^2_a\,T\ e^{\mu_a (T)/T}}{2\pi^2}\
K_2\bigg(\frac{m_a}{T}\bigg)\; ,
\end{eqnarray}
where $\mu_a$ is the  $T$-dependent chemical potential and $K_n(x)$ is
the  $n$th-order modified Bessel  function~\cite{AS}.  In  our minimal
leptogenesis model, the $g_a$ factors are: $g_{W^a} = 3 g_{B} = 6$ and
$g_\Phi = g_{\Phi^\dagger} = 2$,  and for the $i$th family: $g_{N_i} =
2$,  $g_{L_i}  = g_{L^C_i}  =  4$, $g_{Q_i}  =  g_{Q^C_i}  = 12$,  and
$g_{u_i}  =  g_{u_i^C}  =  6$.   If  the  chemical  potential  $\mu_a$
vanishes,  i.e.\ $\mu_a  = 0$,  $n_a (T)$  automatically  satisfies an
in-equilibrium number-density  distribution, which is  usually denoted
as~\cite{KW,MAL} $n^{\rm eq}_a  (T)$ and takes on the  simple forms in
certain limits:
\begin{equation}
  \label{neq}
n^{\rm eq}_a (T)\ =\ \left\{\begin{array}{cc}
g_a \Big(\frac{\displaystyle m_a T}{\displaystyle2\pi}\Big)^{3/2}\, 
e^{-m_a/T}\,,\qquad & (m_a \gg T);\\[2mm]
\frac{\displaystyle g_a\,T^3}{\displaystyle \pi^2}\ ,\qquad & 
(m_a \ll T)\, .\end{array} \right.
\end{equation}
Here,  we   should  note  that  although  the   condition  of  thermal
equilibrium does not imply by itself the vanishing of the chemical
potential,  the in-equilibrium number  density $n^{\rm  eq}_a$ defined
for $\mu_a = 0$ is a useful quantity in writing the BEs later on.

In  analogy to the  formalism introduced in~\cite{MAL},  let us define
the CP-conserving collision  term for a  generic process $X\to  Y$ and
its CP-conjugate one $\overline{X} \to \overline{Y}$ as
\begin{equation}
  \label{CT}
\gamma^X_Y\ \equiv \ \gamma ( X\to Y)\: +\: \gamma ( \overline{X}
\to \overline{Y} )\; ,
\end{equation}
with
\begin{equation}
  \label{gamma}
\gamma ( X\to Y)\ =\ \int\! d\pi_X\, d\pi_Y\, (2\pi )^4\,
\delta^{(4)} ( p_X - p_Y )\ e^{-p^0_X/T}\, |{\cal M}( X \to Y )|^2\; .
\end{equation}
In the above, $|{\cal M}( X \to Y )|^2$  is the squared matrix element
which is summed but  {\em not} averaged over  the internal  degrees of
freedom of the initial and final multiparticle states $X$ and $Y$.  In
addition, we have used the   short-hand notation for the   phase-space
factors:
\begin{equation}
  \label{dpiX}
d\pi_X\ =\ \frac{1}{S_X}\, \prod\limits_{i=1}^{n_X}\,
\frac{d^4 p_i}{(2\pi )^3}\ \delta ( p^2_i - m^2_i )\; \theta (p^0_i)\; ,
\end{equation}
where $S_X = n_{\rm id}!$ is  a symmetry factor in case $X$ contains a
number $n_{\rm  id}$ of identical particles.   An analogous definition
holds for $d\pi_Y$  related to the final multiparticle  state $Y$. 
Since CPT is preserved, the CP-conserving collision term $\gamma^X_Y$
obeys the relation
\begin{equation}
  \label{CPTrel}
\gamma^X_Y \ =\ \gamma^Y_X\; .
\end{equation}
In addition  to the CP-conserving collision  term $\gamma^X_Y$, we may
analogously define a CP-violating collision term as
\begin{equation}
  \label{dgamma}
\delta \gamma^X_Y\ \equiv\ \gamma (X\to Y)\ -\ \gamma (\overline{X}
\to \overline{Y})\ 
=\ -\,\delta \gamma^Y_X\; ,
\end{equation}
where the last equality in~(\ref{dgamma}) follows from CPT invariance.

Following~\cite{KW,MAL}, the BEs for the number densities $n_a$ of all
particle  species  $a$  in  a  given  model  form  a  set  of  coupled
first-order   differential  equations.   These   coupled  differential
equations can generically be written down as
\begin{equation}
  \label{BEgeneric}
\frac{dn_a}{dt}\: +\: 3 H n_a\ =\ -\, \sum\limits_{aX^\prime\leftrightarrow
  Y}\,\bigg[\ \frac{n_a n_{X^\prime}}{n^{\rm eq}_a 
  n^{\rm eq}_{X^\prime}}\,\gamma (a X^\prime \to Y)\ -\ 
\frac{n_Y}{n^{\rm eq}_Y}\, \gamma (Y\to a X^\prime )\ \bigg]\; ,
\end{equation}
where the sum is over all possible reactions in which the particle $a$
can  be annihilated  or  created through  a  reaction of  the form  $a
X^\prime \to Y$ or  $Y\to a X^\prime$.  Special treatment~\cite{KW} is
required if  a particle species $a$  is unstable and  hence allowed to
occur as  a RIS in a  resonant process like  $X \to a\to Y$.   We will
discuss  below  our approach  to  this problem  for  the  case of  the
unstable heavy Majorana neutrinos $N_i$.

In order to reduce the large number of the coupled BEs, we assume that
all chemical potentials of the Higgs field $\Phi$,  the quarks and the
gauge fields $V_\mu$ are significantly smaller than the one associated
with the lepton number $L$.   This assumption can  be justified from a
thermal   equilibrium  analysis of the     chemical potentials.  For a
three-generation model, such an analysis yields~\cite{HT,APreview}:
\begin{equation}
  \label{mupot}
\mu_V\ =\  0\,,\quad  \mu_\Phi\ =\  \frac{4}{21}\,\mu_L\,,\quad  
\mu_Q \ =\ -\,  \frac{1}{3}\,\mu_L\,,\quad  \mu_u\ =\  \frac{5}{21}\,
\mu_L\,, 
\end{equation}
where  $\mu_L  =   \sum_{i=1}^3   \mu_{L_i}$,   $\mu_Q =  \sum_{i=1}^3
\mu_{Q_i}$ and $\mu_u   =  \sum_{i=1}^3  \mu_{u_i}$.   Note that   the
remaining chemical potentials  $\mu_d$ and $\mu_e$, albeit  comparable
to $\mu_L$, enter the  BEs only through sub-dominant  collision terms,
e.g.~the $b$-quark Yukawa-coupling contribution is smaller at least by
a  factor   $10^{-3}$ with respect to   the  corresponding one  due to
$t$-quarks.

Employing   (\ref{BEgeneric}) and    the definitions~(\ref{gamma}) and
(\ref{dgamma}) for the collision  terms, it is now straightforward  to
derive the BEs for $n_{N_i}$ and $n_L$  that govern the time evolution
of the number densities of the heavy Majorana neutrinos and the lepton
number,  respectively.   To   leading order   in  the  small parameter
$n_L/n^{\rm eq}_l$~\footnote{The quantity  $n^{\rm  eq}_l = 2n_\gamma$
is the in-equilibrium number density  for an individual lepton doublet
$L_i$ with $\mu_{L_i} = 0$ [cf.\ (\ref{neq})].}, we obtain
\begin{eqnarray}
  \label{BE1}
\frac{dn_{N_i}}{dt}\: +\: 3 H n_{N_i} \!\!&=&\!\! \bigg( 1 \: -\: 
\frac{n_{N_i}}{n^{\rm eq}_{N_i}}\,\bigg)\, \bigg( 
\gamma^{N_i}_{L\Phi}\: +\: \gamma^{N_i L}_{Q u^C}\: +\: 
\gamma^{N_i u^C}_{L Q^C}\: +\: \gamma^{N_i Q}_{L u}\nonumber\\ 
\!\!&&\!\!+\, \gamma^{N_i V_\mu}_{L\Phi}  \: +\: 
\gamma^{N_i L}_{\Phi^\dagger V_\mu}\: +\: 
\gamma^{N_i\Phi^\dagger }_{LV_\mu}\, \bigg)\nonumber\\
\!\!&&\!\!-\, \frac{n_L}{2n^{\rm eq}_l}\, \bigg[\,
\delta\gamma^{N_i}_{L\Phi}\: +\: 
\delta\gamma^{N_i u^C}_{L Q^C}\: +\: \delta\gamma^{N_i Q}_{L u}\:
+\: \delta\gamma^{N_i V_\mu}_{L\Phi}\: +\: 
\delta\gamma^{N_i\Phi^\dagger }_{LV_\mu}\nonumber\\
\!\!&&\!\! +\: \frac{n_{N_i}}{n^{\rm eq}_{N_i}}\,
\bigg(\, \delta\gamma^{N_i L}_{Q u^C}\: +\: 
\delta\gamma^{N_i L}_{\Phi^\dagger V_\mu}\,\bigg)\,\bigg]\; ,\\[3mm]
  \label{BE2}
\frac{dn_L}{dt}\: +\: 3 H n_L \!\!&=&\!\! -\,\bigg( 1 \: -\: 
\frac{n_{N_i}}{n^{\rm eq}_{N_i}}\,\bigg)\, \bigg( 
\delta\gamma^{N_i}_{L\Phi}\: -\: \delta\gamma^{N_i L}_{Q u^C}\: +\: 
\delta\gamma^{N_i u^C}_{L Q^C}\: +\: \delta\gamma^{N_i Q}_{L u}\nonumber\\ 
\!\!&&\!\!+\, \delta\gamma^{N_i V_\mu}_{L\Phi}\: -\: 
\delta\gamma^{N_i L}_{\Phi^\dagger V_\mu}\: +\: 
\delta\gamma^{N_i\Phi^\dagger }_{LV_\mu}\, \bigg)\nonumber\\
\!\!&&\!\!-\, \frac{n_L}{2n^{\rm eq}_l}\, \bigg[\,
\gamma^{N_i}_{L\Phi}\: +\: 2\gamma^{\,\prime L\Phi}_{\,L^C\Phi^\dagger} 
+\:  4\gamma^{LL}_{\Phi^\dagger\Phi^\dagger} +\: 
2\gamma^{N_i L}_{Q u^C} +\:
2\gamma^{N_i u^C}_{L Q^C}\, +\: 2\gamma^{N_i Q}_{L u}\nonumber\\
&&+\: 2\gamma^{N_i V_\mu}_{L\Phi} +\:
2\gamma^{N_i L}_{\Phi^\dagger V_\mu}  +\:  
2\gamma^{N_i\Phi^\dagger }_{LV_\mu} +\: 
\frac{n_{N_i}}{n^{\rm eq}_{N_i}}\, \bigg(\, \gamma^{N_i L}_{Q u^C}\: +\: 
\gamma^{N_i L}_{\Phi^\dagger V_\mu}\,\bigg)\,\bigg]\; .\quad
\end{eqnarray}
In~(\ref{BE2}), $\gamma^{\,\prime L\Phi}_{\,L^C\Phi^\dagger}$  denotes
the    collision  term   defined   in~(\ref{dgamma})   after   the RIS
contributions due    to heavy  Majorana  neutrinos   $N_i$  have  been
subtracted  (see also our  discussions in Section~\ref{sec:RIS} and in
Appendix~\ref{app:CT}).

In  order to  terminate the  infinite series  of collision  terms that
could  be  added  in  the  BEs~(\ref{BE1})  and~(\ref{BE2}),  we  have
developed a  systematic expansion in powers of  coupling constants for
all     $1\leftrightarrow      2$     and     $2\leftrightarrow     2$
processes~\footnote{As   we  will   see   in  Section~\ref{sec:BEs}.3,
$1\leftrightarrow     2$     processes     become     important     at
$T\stackrel{<}{{}_\sim}   m_{N_1}$,   whereas   $2\leftrightarrow   2$
scatterings   dominate   at   $T\stackrel{>}{{}_\sim}  m_{N_1}$   (see
also~\cite{BCST,BBP}).  In this respect,  $1 \leftrightarrow 3$ and $2
\leftrightarrow  3$  processes  should  be  regarded  as  higher-order
corrections   to   the    corresponding   $1\leftrightarrow   2$   and
$2\leftrightarrow  2$  processes.}.   More  precisely, for  all  these
processes  that involve  only  one heavy  Majorana  neutrino, we  have
included all  collision terms depending  on the coupling  constants as
$(\bar{h}_\pm^\nu)^2$, $(\bar{h}_\pm^\nu)^2 g^2$, $(\bar{h}_\pm^\nu)^2
g^{\prime 2}$ and  $(\bar{h}^\nu_\pm)^2 h^2_u$, where $\bar{h}^\nu_\pm
\sim  h^\nu$  are the  one-loop  resummed  effective Yukawa  couplings
calculated   in   Section~\ref{sec:RIS}.    In  fact,   we   neglected
$2\leftrightarrow  2$  scatterings  ${\cal  O} [(h^\nu)^4]$  with  two
external heavy  Majorana neutrinos,  such as $N_i  N_j \leftrightarrow
LL$  and its CP-conjugate  part.  Instead,  we included  the collision
terms of order $(h^\nu)^4$ for $2 \leftrightarrow 2$ scatterings where
all   external  particles   are  massless,   e.g.~$LL  \leftrightarrow
\Phi^\dagger\Phi^\dagger$, $L\Phi\leftrightarrow L^C\Phi^\dagger$.

An important  intermediate step in the  derivation of  (\ref{BE2}) has
been  the proper implementation of the  relations of unitarity and CPT
invariance  that  govern the collision  terms  pertaining to reactions
with  different number   of external   particles,   i.e.\ between  the
RIS-subtracted $2\leftrightarrow 2$ scatterings and  $1\leftrightarrow
2$ processes or   between   the RIS-subtracted   $3\leftrightarrow  2$
processes and  $2\leftrightarrow  2$  reactions.  More  explicitly, on
account of  unitarity  and CPT invariance,  the following perturbative
relations for  the CP-violating parts  of the RIS-subtracted collision
terms can be established
\begin{eqnarray}
  \label{CPT1}
\gamma\,'(L\Phi \to L^C\Phi^\dagger)\: -\: 
\gamma\,'(L^C\Phi^\dagger \to L\Phi) 
&=& \delta\gamma^{N_i}_{L\Phi}\ +\ {\cal O}[(h^\nu)^4]\;,\nonumber\\
\gamma\,'(L Q^C \to L^C \Phi^\dagger u^c)\: -\: 
\gamma\,'(L^C Q \to L \Phi u ) &=&  \delta\gamma^{N_i u^C}_{L Q^C}\ 
+\ {\cal O}[(h^\nu)^4 h^2_u]\;,\nonumber\\
\gamma\,' (Q u^C \to L L \Phi )\: -\: 
\gamma\,' ( Q^C u \to L^C L^C \Phi^\dagger ) &=& 
\delta\gamma^{N_i L}_{Qu^C}\ +\ {\cal O}[(h^\nu)^4
  h^2_u]\qquad {\rm etc.},\quad
\end{eqnarray}
where the prime defines an operation of  RIS subtraction.  Notice that
the omission  of the higher-order terms  is fully consistent  with our
truncated expansion outlined above.  In this context, we also formally
neglected  as higher-order  effects  the  CP-conserving RIS-subtracted
collision terms related to  $2\to 3$ scatterings.   

Unlike $2\to 3$ scatterings, $3 \to 2$ scatterings, e.g.~$L \Phi u \to
L^C Q$, $L  L \Phi \to Qu^C$  etc., should not  be subtracted, as they
have not been counted before.   Moreover, CPT invariance and unitarity
give rise to the following constraints for this set of reactions:
\begin{eqnarray}
  \label{CPT2}
\gamma (L \Phi u \to L^C Q  )\: -\: 
\gamma (L^C \Phi^\dagger u^C \to L Q^C )
 &=& {\cal O}[(h^\nu)^4 h^2_u]\;,\nonumber\\
\gamma (L L \Phi \to Q u^C )\: -\: 
\gamma ( L^C L^C \Phi^\dagger \to Q^C u ) &=& {\cal O}[(h^\nu)^4
  h^2_u]\qquad {\rm etc.}\quad
\end{eqnarray}
As a consequence,  the   $3\to 2$ scatterings contribute    additional
CP-conserving wash-out  $2\leftrightarrow 2$  scattering terms to  the
BE~(\ref{BE2}),  through  the   resonant exchange  of  heavy  Majorana
neutrinos $N_i$. These extra wash-out  terms can easily be  calculated
by  applying    the  narrow-width  approximation to   the    $3\to  2$
scatterings, e.g.
\begin{eqnarray}
  \label{NWA}
\gamma (L^C \Phi^\dagger u^C \to L Q^C )\: +\: 
\gamma (L \Phi u \to L^C Q  )
 &=& \frac{1}{2}\, \gamma^{N_i u^C}_{L Q^C}\: 
+\: {\cal O}[(h^\nu)^4 h^2_u]\;,\nonumber\\
\gamma (L L \Phi \to Q u^C )\: +\: 
\gamma ( L^C L^C \Phi^\dagger \to Q^C u ) &=& 
\frac{1}{2}\,\gamma^{N_i L}_{Qu^C}\: +\: {\cal O}[(h^\nu)^4
  h^2_u]\qquad {\rm etc.}\quad
\end{eqnarray}
Note  that these last relations have  already  been implemented in the
BE~(\ref{BE2}).  Finally, since CP violation is predominantly mediated
by the    resonant exchange of  heavy   Majorana neutrinos  $N_i$, the
CP-violating collision terms can be  further approximated in terms  of
the CP-conserving ones as follows:
\begin{eqnarray}
  \label{simpl}
\delta \gamma^{N_i}_{L\Phi} \!\!&=&\!\! 
             \delta_{N_i}\, \gamma^{N_i}_{L\Phi}\,,\qquad
\delta\gamma^{N_i u^C}_{L Q^C}\ =\ 
             \delta_{N_i}\,\gamma^{N_i u^C}_{L Q^C}\,,\qquad
\delta\gamma^{N_i L}_{Qu^C}\ =\ -\,
             \delta_{N_i}\,\gamma^{N_i L}_{Qu^C}\qquad {\rm etc.},\quad
\end{eqnarray}
where the CP  asymmetries $\delta_{N_i}$ are  given in~(\ref{deltaNi})
and   all   CP-conserving    collision    terms  are   presented    in
Appendix~\ref{app:CT}.

To numerically solve the BE's, we introduce a number of new variables.
This will also  enable us to  compare our results with the literature.
To this end, we make  use of the relation between  the cosmic time $t$
and the temperature~$T$:
\begin{equation}
  \label{Tt}
t \ =\ \frac{z^2}{2\, H(z=1)}\ ,
\end{equation}
where
\begin{equation}
  \label{zeta}
z\ =\ \frac{m_{N_1}}{T}\ .
\end{equation}
The  relation~(\ref{Tt}) is valid in  the radiation-dominated epoch of
the Universe relevant to  baryogenesis. In addition, we introduce  the
parameters  $\eta_a$  which  give  the  number density of  a  particle
species $a$ normalized to the number density of photons, i.e.
\begin{equation}
  \label{etas}
\eta_a (z) \ =\ \frac{n_a (z)}{n_\gamma (z)}\ , 
\end{equation}
with
\begin{equation}
  \label{ngamma}
n_\gamma (z)\ =\ \frac{2\,T^3}{\pi^2}\ =\ 
               \frac{2\, m^3_{N_1}}{\pi^2}\,\frac{1}{z^3}\ .
\end{equation}
With the above definitions, the BEs~(\ref{BE1}) and (\ref{BE2}) can be
written down in a more compact form:
\begin{eqnarray}
  \label{BEN} 
\frac{d \eta_{N_i}}{dz} &=& \frac{z}{H(z=1)}\ \bigg[\,\bigg( 1
\: -\: \frac{\eta_{N_i}}{\eta^{\rm eq}_{N_i}}\,\bigg)\, \bigg(\,
\Gamma^{D\; (i)} \: +\: \Gamma^{S\; (i)}_{\rm Yukawa}\: +\:
\Gamma^{S\; (i)}_{\rm Gauge}\, \bigg) \nonumber\\ 
&&-\, \frac{1}{4}\, \eta_L\, \delta_{N_i}\, \bigg(\, \Gamma^{D\; (i)} \: +\:
\widetilde{\Gamma}^{S\; (i)}_{\rm Yukawa}\: +\:
\widetilde{\Gamma}^{S\; (i)}_{\rm Gauge}\, \bigg)\,\bigg]\,,\\[3mm] 
  \label{BEL}
\frac{d \eta_L}{dz} &=& -\, \frac{z}{H(z=1)}\, \bigg\{\,
\sum\limits_{i=1}^3\, \delta_{N_i}\ 
\bigg( 1 \: -\: \frac{\eta_{N_i}}{\eta^{\rm eq}_{N_i}}\,\bigg)\, \bigg(\,
\Gamma^{D\; (i)} \: +\: \Gamma^{S\; (i)}_{\rm Yukawa}\: +\:
\Gamma^{S\; (i)}_{\rm Gauge}\, \bigg) \nonumber\\ 
&&+\, \frac{1}{4}\, \eta_L\, \bigg[\, \sum\limits_{i=1}^3\, 
\bigg(\, \Gamma^{D\; (i)} \: +\: 
\Gamma^{W\;(i)}_{\rm Yukawa}\: 
+\: \Gamma^{W\; (i)}_{\rm Gauge}\,\bigg)\: +\:
\Gamma^{\Delta L =2}_{\rm Yukawa} \bigg]\,\bigg\}\,,
\end{eqnarray}
where 
\begin{eqnarray}
  \label{GD}
\Gamma^{D\; (i)} & = & \frac{1}{n_\gamma}\ \gamma^{N_i}_{L\Phi}\;, \nonumber\\
  \label{GSY}
\Gamma^{S\; (i)}_{\rm Yukawa} & = & \frac{1}{n_\gamma}\
\bigg(\, \gamma^{N_i L}_{Q u^C}\: +\:  \gamma^{N_i u^C}_{L Q^C}\: 
+\: \gamma^{N_i Q}_{L u}\, \bigg)\; ,\nonumber\\
\widetilde{\Gamma}^{S\;(i)}_{\rm Yukawa} &=& \frac{1}{n_\gamma}\
\bigg(\, \frac{\eta_{N_i}}{\eta^{\rm eq}_{N_i}}\, \gamma^{N_i L}_{Q u^C}\: 
+\: \gamma^{N_i u^C}_{L Q^C}\: +\: \gamma^{N_i Q}_{L u}\, \bigg)\;,\nonumber\\
  \label{GSG}
\Gamma^{S\; (i)}_{\rm Gauge} & = & \frac{1}{n_\gamma}\ 
\bigg(\, \gamma^{N_i V_\mu}_{L\Phi}\: +\: 
\gamma^{N_i L}_{\Phi^\dagger V_\mu}\: +\: 
\gamma^{N_i\Phi^\dagger }_{LV_\mu}\, \bigg)\;,\nonumber\\
\widetilde{\Gamma}^{S\; (i)}_{\rm Gauge} &=& \frac{1}{n_\gamma}\ 
\bigg(\, \gamma^{N_i V_\mu}_{L\Phi}\: +\: 
\frac{\eta_{N_i}}{\eta^{\rm eq}_{N_i}}\, 
\gamma^{N_i L}_{\Phi^\dagger V_\mu}\: +\: 
\gamma^{N_i\Phi^\dagger }_{LV_\mu}\, \bigg)\; ,\nonumber\\
  \label{GWY}
\Gamma^{W\; (i)}_{\rm Yukawa} & = & \frac{2}{n_\gamma}\
\bigg(\, \gamma^{N_i L}_{Q u^C}\: +\:  \gamma^{N_i u^C}_{L Q^C}\: 
+\: \gamma^{N_i Q}_{L u}\: +\: \frac{\eta_{N_i}}{2\eta^{\rm eq}_{N_i}}\,
\gamma^{N_i L}_{Q u^C}\, \bigg)\; ,\nonumber\\
  \label{GWG}
\Gamma^{W\; (i)}_{\rm Gauge} & = & \frac{2}{n_\gamma}\ 
\bigg(\, \gamma^{N_i V_\mu}_{L\Phi}\: +\: 
\gamma^{N_i L}_{\Phi^\dagger V_\mu}\: +\: 
\gamma^{N_i\Phi^\dagger }_{LV_\mu}\: +\:
\frac{\eta_{N_i}}{2\eta^{\rm eq}_{N_i}}\, 
\gamma^{N_i L}_{\Phi^\dagger V_\mu}\, \bigg)\;,\nonumber\\
  \label{GDL2}
\Gamma^{\Delta L =2}_{\rm Yukawa} &=& \frac{2}{n_\gamma}\ 
\bigg(\, \gamma^{\,\prime L\Phi}_{\,L^C\Phi^\dagger} 
+\:  2\gamma^{LL}_{\Phi^\dagger\Phi^\dagger}\, \bigg)\; .
\end{eqnarray}
In   writing  the   BEs~(\ref{BEN})  and~(\ref{BEL}),   we   used  the
approximate    relations~(\ref{simpl}).   Hence,    all   CP-violating
contributions do only depend on the parameters $\delta_{N_{1,2,3}}$.

Finally,      it  is  worth    stressing     that the  BEs~(\ref{BEN})
and~(\ref{BEL}) can be applied without any additional restriction to a
more  general context of thermal  leptogenesis, including scenarios of
non-resonant  leptogenesis.   Most    importantly,  we observe    that
CP-violating scatterings provide  an additional  non-negligible source
of CP  violation and may lead  to an increase  in the predicted values
for  $\eta_L$ at    $T\stackrel{>}{{}_\sim} m_{N_1}$,  as opposed   to
previous studies where those   effects  were neglected.  In the   next
section, we will present numerical estimates  for the generated BAU in
a few representative  light-neutrino models compatible with  solar and
atmospheric neutrino data.

\subsection{Numerical Examples}

In  our numerical analysis, we will  consider scenarios with two nearly
degenerate heavy Majorana neutrinos  $N_{1,2}$ with masses at the  TeV
range.   The mass of the third  heavy  Majorana neutrino $N_3$ will be
taken to be of order  $10^{15}$~GeV, so the  decoupling of $N_3$  from
the low-energy  sector of the  theory is   natural.  For  the  generic
light-neutrino model described by~(\ref{mnuappr}), the neutrino Yukawa
couplings $(h^\nu)_{l1,2}$ are expressed in the mass basis as
\begin{equation}
  \label{hnumodel}
h^\nu_{l1}\ =\ \frac{1}{\sqrt{2}}\; \Big( h_{l1}\,\varepsilon\: 
+\: h_{l2}\,\bar{\varepsilon}\,\Big)\,,\qquad 
h^\nu_{l2}\ =\ \frac{i}{\sqrt{2}}\; \Big( h_{l1}\,\varepsilon\: 
-\: h_{l2}\,\bar{\varepsilon}\,\Big)\,,\qquad 
\end{equation}
where the Yukawa  couplings $h_{l1,2}$ are given in~(\ref{hmodel}). In
addition,     the  FN    expansion     parameters   $\varepsilon$  and
$\bar{\varepsilon}$,   or equivalently   $h_{l1}$  and  $h_{l2}$,  are
assumed to    be  complex.     For  the  models    of  our   interest,
$|\varepsilon|$ and $|\bar{\varepsilon}|$ are taken to be smaller than
$10^{-3}$.  According to  our discussion  in Section~\ref{sec:models},
the  scenario given by~(\ref{hnumodel})  can  accommodate the neutrino
data,  provided  $|\varepsilon\,\bar{\varepsilon}| \sim m_{N_1}/M_{\rm
GUT}$. Therefore,  in  addition to scenarios with   $|\varepsilon| \sim
|\bar{\varepsilon}|$, we will also present  numerical estimates of the
BAU  for models with   $|\bar{\varepsilon}| \ll |\varepsilon |$, where
the  product   $|\varepsilon\,\bar{\varepsilon}|$   is   fixed      to
$m_{N_1}/M_{\rm GUT}$.

\begin{figure}
  \begin{center}
   \setlength{\unitlength}{0.90cm}
    \begin{picture}(13.5,20)(0,0)
      \Text(-0.7,15.7)[]{\bf (a)}
      \Text(-0.7,5.6)[]{\bf (b)}
      \put(0,0){\includegraphics{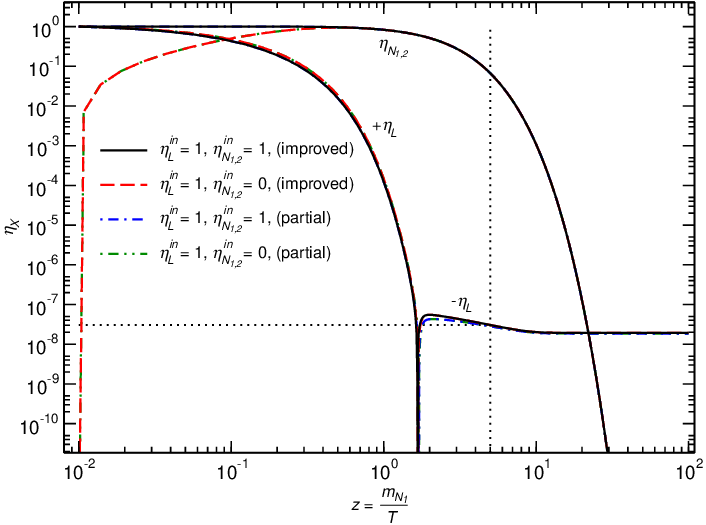}}
      \put(0,10.2){\includegraphics{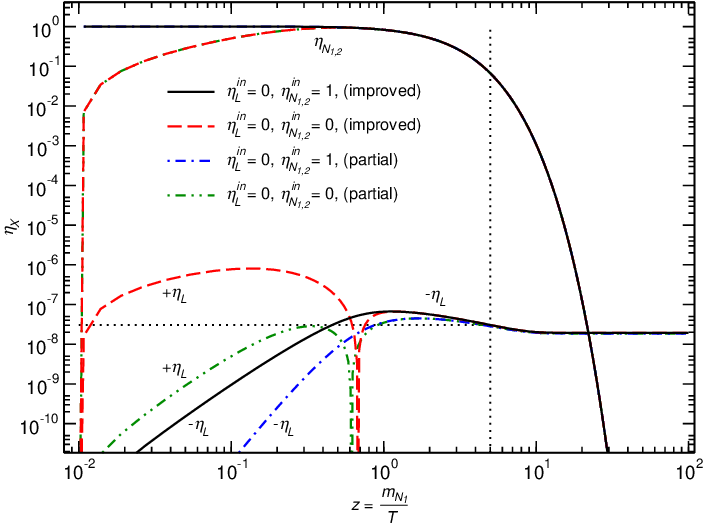}}
    \end{picture}
  \end{center}
\vspace{-0.5cm}
\caption{\em  Numerical  estimates  of $\eta_L$,  $\eta_{N_{1,2}}$  as
functions of $z = m_{N_1}/T$,  for a model where $m_{N_1}=1$~TeV, $x_N
= \frac{m_{N_2}}{m_{N_1}}-1=   7.7\times 10^{-10}$,  $\varepsilon    =
4.3\times10^{-7}$,  $\bar{\varepsilon}= -i\,4.3\times10^{-7}$, and for
{\bf         (a)}~$\eta^{\,\mathrm{in}}_{L}=0$        and         {\bf
(b)}~$\eta^{\,\mathrm{in}}_{L}=1$.   The horizontal  dotted line shows
the value  of $\eta_L$ needed to  produce the observed  $\eta_B$.  The
vertical dotted line corresponds to $T = T_c = 200$~GeV.  ``Improved''
and  ``partial'' refer to whether or   not the CP-violating scattering
terms proportional to  $\delta_{N_i}\left(\Gamma^{S\;(i)}_{\rm Yukawa}
+    \Gamma^{S\;(i)}_{\rm    Gauge}\right)$  are     included  in  the
BE~(\ref{BEL}).}\label{fig:num1} \setlength{\unitlength}{1pt}
\end{figure}

We start our  numerical analysis by  exhibiting in Fig.~\ref{fig:num1}
numerical  values    of the   lepton     asymmetry $\eta_L$   and  the
heavy-neutrino number  densities $\eta_{N_{1,2}}$ as functions  of the
parameter $z = m_{N_1}/T$.  Specifically, we have set $m_{N_1}=1$~TeV,
$x_N = \frac{m_{N_2}}{m_{N_1}}-1=  7.7\times 10^{-10}$, $\varepsilon =
4.3\times10^{-7}$,  $\bar{\varepsilon}  = -i\,4.3\times10^{-7}$.   The
horizontal  dotted line    shows the  value  of   $\eta_L$  needed  at
temperatures $T$ close  to  the critical  temperature $T_c =  200$~GeV
(indicated by  the   vertical dotted line)   to  produce the  observed
$\eta_B$.       In  Fig.~\ref{fig:num1}(a),     we        have   taken
$\eta^{\,\mathrm{in}}_{L}=0$  as the   initial value   of the leptonic
asymmetry.  As  initial   conditions  for the    heavy-neutrino number
densities  $\eta_{N_{1,2}}$, we   have considered  two  possibilities,
depending  on whether the heavy    neutrinos are initially in  thermal
equilibrium,     $\eta^{\rm   in}_{N_{1,2}}  =    1$,    or   strongly
out-of-equilibrium, $\eta^{\rm in}_{N_{1,2}} = 0$.  On the same panel,
we also show numerical results obtained if the CP-violating scattering
terms proportional to $\delta_{N_i}\left(\Gamma^{S\;(i)}_{\rm  Yukawa}
+  \Gamma^{S\;(i)}_{\rm  Gauge}\right)$   are  not  included  in   the
BE~(\ref{BEL}).  The latter  is the approach  followed in the existing
literature.  We refer  to such a  numerical computation as  `partial'.
For the model under discussion,  these results are compared with those
obtained with our `improved' treatment where those terms are included.
We find  that the  two predictions  for  $\eta_B$ at  $T  \ll m_{N_1}$
($z\gg 1$) turn out  to come very close together;  they only differ by
$\sim 3\%$.  However, at $T \sim m_{N_{1,2}}$ corresponding to $z \sim
1$, the `improved'  and `partial'  computations  may differ even  by a
factor of 3.  Most interestingly, with the new CP-violating scattering
terms included,  the   asymptotic value   of  the leptonic   asymmetry
$\eta_L$   is attained at a  somewhat  higher temperature  than in the
`partial' approach.

An    important consequence  of   resonant  leptogenesis  is that  the
generated  leptonic asymmetry   $\eta_L(T)$   is independent of    the
primordial  (initial) lepton asymmetry   $\eta^{\rm in}_L$.  As can be
seen from   Fig.~\ref{fig:num1}(b),  even   if  the  initial  leptonic
asymmetry is  taken to be maximal  corresponding to $\eta^{\rm in}_L =
1$,  this primordial lepton  asymmetry   gets rapidly  erased and  the
predicted   $\eta_L$ is  finally  set  by  the resonant   leptogenesis
mechanism itself for   $z\stackrel{>}{{}_\sim} 1$.   Because of   the
independence of $\eta_L$ at $T=T_*\sim T_c$ on the initial conditions,
in  the   following we will   only  show  numerical  values  using our
`improved' approach to BEs for  initial conditions: $\eta^{\rm in}_L =
0$ and $\eta^{\rm in}_{N_{1,2}} = 1$.

\begin{figure}
  \begin{center}
   \setlength{\unitlength}{0.90cm}
    \begin{picture}(13.5,20)(0,0)
      \Text(-0.7,15.7)[]{\bf (a)}
      \Text(-0.7,5.6)[]{\bf (b)}
      \put(0,0){\includegraphics{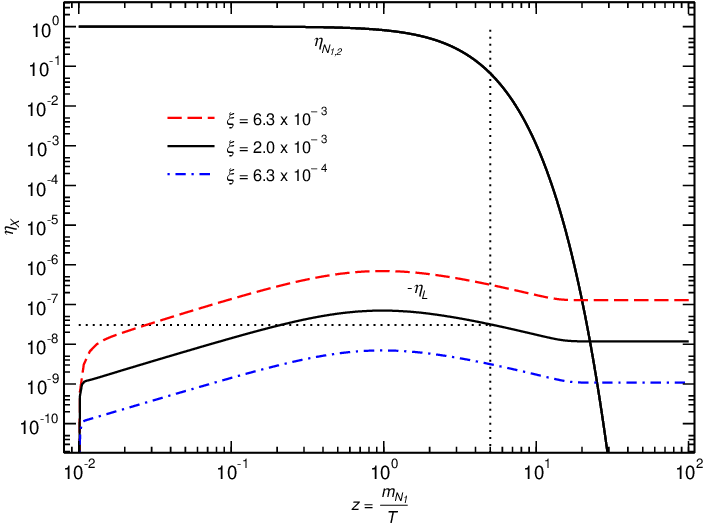}} 
      \put(0,10.2){\includegraphics{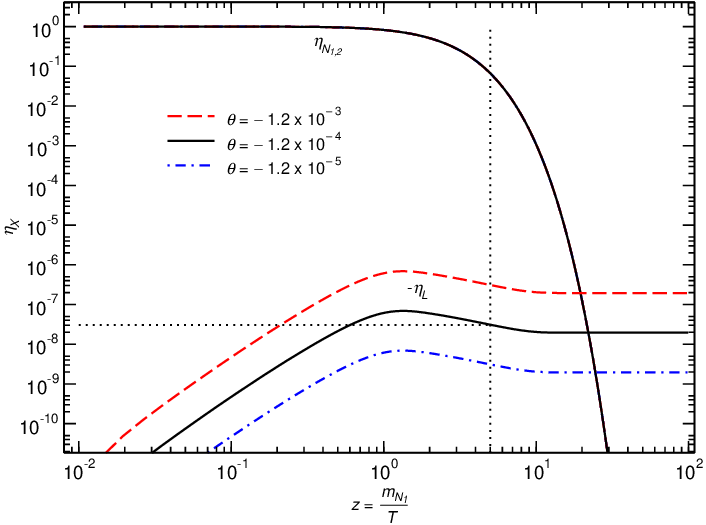}}
    \end{picture}
  \end{center}
\vspace{-0.5cm}
\caption{\em  Numerical  estimates  of $\eta_L$,  $\eta_{N_{1,2}}$  as
functions of $z =  m_{N_1}/T$, for two scenarios with $m_{N_1}=1$~TeV,
$x_N  = \frac{m_{N_2}}{m_{N_1}}  - 1  = \varepsilon^2$  and $\eta^{\rm
in}_L   =    0$,   $\eta^{\rm   in}_{N_{1,2}}   =    1$:   {\bf   (a)}
$\bar{\varepsilon}  =  e^{i\theta}  \varepsilon$  and  $\varepsilon  =
4.3\times10^{-7}$;  {\bf (b)}  $\bar{\varepsilon} =  i\xi \varepsilon$
and  $|\varepsilon  \bar{\varepsilon}|  = 1.85  \times10^{-13}$.   The
meaning of the horizontal and vertical  dotted lines is the same as in
Fig.~\ref{fig:num1}.}\label{fig:num2} \setlength{\unitlength}{1pt}
\end{figure}

We now present numerical predictions of the BAU for two representative
variants  of our  generic  model,  where we  implement the  additional
constraint, $x_N =  \frac{m_{N_2}}{m_{N_1}} - 1 = \varepsilon^2$ (with
$\varepsilon$   being   real),  according   to    our   discussion  in
Section~\ref{sec:models}.   Fig.~\ref{fig:num2} shows numerical values
of $\eta_L$, $\eta_{N_{1,2}}$ as functions of $z = m_{N_1}/T$, for two
scenarios with $m_{N_1}=1$~TeV  and $\eta^{\rm in}_L = 0$,  $\eta^{\rm
in}_{N_{1,2}} =  1$: (a)~$\bar{\varepsilon} = e^{i\theta} \varepsilon$
and $\varepsilon =   4.3\times10^{-7}$; (b)~$\bar{\varepsilon} =  i\xi
\varepsilon$ (with $\xi  < 1$) and  $|\varepsilon \bar{\varepsilon}| =
1.85  \times10^{-13}$.   As before,  the   adopted values for  the  FN
expansion  parameters $\varepsilon$ and $\bar{\varepsilon}$ are chosen
to be  in agreement with neutrino  data.   In the first  variant where
$|\bar{\varepsilon}| = |\varepsilon  |$, we see from~\ref{fig:num2}(a)
that  CP-violating phases  $|\theta  |$  as   small as $10^{-4}$   are
sufficient to   account for the  observed BAU.    In  this small phase
regime, the predicted values for $\eta_L$ and consequently for the BAU
scale linearly with the CP-violating phase $\theta$.

The second variant of our generic  model realizes a large CP-violating
phase ($\theta =  \pi/2$), but introduces an  hierarchy between the FN
parameters, i.e.~$\bar{\varepsilon} =  i\xi \varepsilon$,  with $\xi <
1$.  {}From~\ref{fig:num2}(b) we   observe that a mild hierarchy  with
values of $\xi$  of order $10^{-3}$--$10^{-2}$   can well explain  the
BAU. As can also be seen  from Fig.~\ref{fig:num2}(b), the predictions
for $\eta_L$  scale quadratically with  the hierarchy factor  $\xi$ in
this scenario.  A particularly     interesting point that    needs  be
emphasized here  is that  the   scenarios with  $\xi \approx   2\times
10^{-3}$ and  $10^{-3}$   give  rise to   out-of-equilibrium-departure
factors $K_i$ defined   in~(\ref{Ki})  much  larger than  $10^3$.   In
particular, we find  that the simple order-of-magnitude estimate given
in~(\ref{etaBapprox}) remains valid for  $K_i$  values at least up  to
order $10^4$.

In Figs.~\ref{fig:num1} and \ref{fig:num2}, we have also displayed the
dependence  of the  heavy-neutrino number  densities $\eta_{N_{1,2}}$,
for  temperatures $T\ll  T_c$, where  the  $(B+L)$-violating sphaleron
interactions  were turned  off.   We see  that  it is  $\eta_{N_{1,2}}
\stackrel{<}{{}_\sim} 10^{-10}$ at  temperatures $T\gg 1$~GeV.  Hence,
the heavy Majorana neutrinos  are under-abundant, much before the epoch
of big-bang nucleosynthesis~\cite{KT}.  This conclusion holds true for
heavy Majorana-neutrino masses  as low as 0.3~TeV.  In  this case, the
observed BAU can still be generated, with the only difference that the
reprocessing of leptons  into baryons will freeze out  at smaller $z$,
i.e.~$z\sim  1$.   In  this   context,  one  may  wonder  whether  the
leptogenesis  scale  can be  lowered  even  further, by  contemplating
models  where the  freeze-out  of  sphalerons happens  at  $z \ll  1$.
Although this  is in principle  possible, one should  expect, however,
that thermal and non-perturbative sphaleron effects will start playing
a crucial role for this class of models and hence a different approach
based  on space-time-dependent diffusion  equations~\cite{EWBAU} would
be more appropriate to reliably address such scenarios.

\begin{figure}[t]
  \begin{center}
   \setlength{\unitlength}{0.90cm}
    \begin{picture}(13.5,10)(0,0)
      \put(0,0){\includegraphics{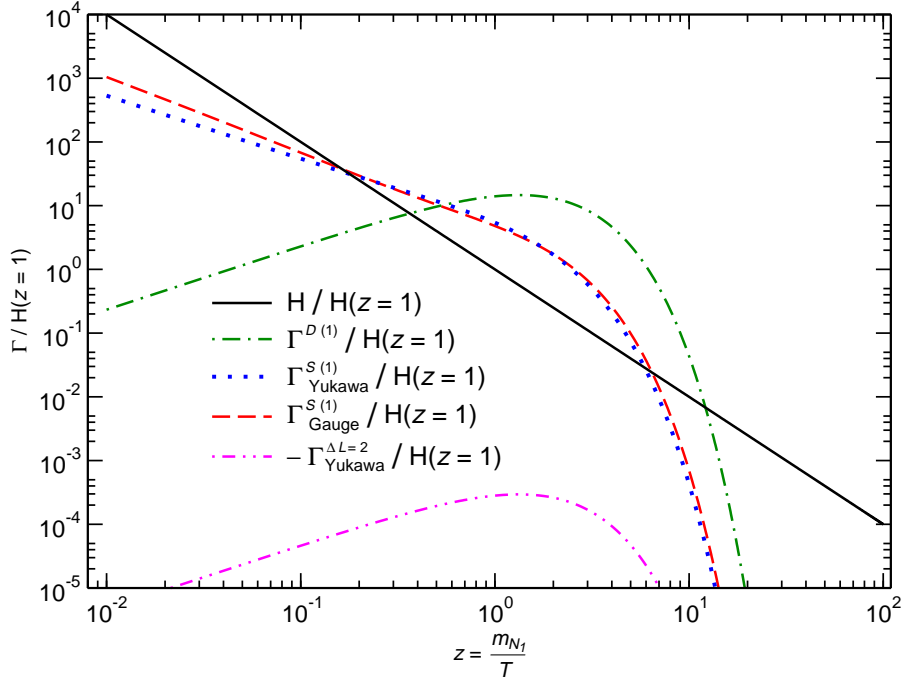}}
    \end{picture}
  \end{center}
\vspace{-0.5cm}
\caption{\em  Functional  dependence of  the  various collision  terms
contributing to the BEs on $z=m_{N_1}/T$. The input parameters are the
same  as those  in Fig.~\ref{fig:num2}(a)  for $\theta  =  - 1.6\times
10^{-4}$.   The   corresponding  wash-out  contributions  $\Gamma^{W\;
(1)}_{\rm Yukawa}$  and $\Gamma^{W\; (1)}_{\rm Gauge}$  are not shown,
as  they  are approximately:  $\Gamma^{W\;  (1)}_{\rm Yukawa}  \approx
2\Gamma^{S\;  (1)}_{\rm  Yukawa}$  and $\Gamma^{W\;  (1)}_{\rm  Gauge}
\approx     2\Gamma^{S\;    (1)}_{\rm     Gauge}$.}\label{fig:num3}
\setlength{\unitlength}{1pt}
\end{figure}

We conclude this  section by commenting on two  points.  First, in our
numerical analysis the gauge-mediated  collision terms provide one  of
the  dominant   sources    of  the   scattering    collision terms  at
$T\stackrel{>}{{}_\sim}   m_{N_1}$.       As  is      illustrated   in
Fig.~\ref{fig:num3},    the  scattering     collision             term
$\Gamma^{S\;(i)}_{\rm   Gauge}$ is comparable to $\Gamma^{S\;(i)}_{\rm
Yukawa}$ which  describes  the top-Yukawa interactions.  Observe  that
$\Gamma^{\Delta L =2}_{\rm Yukawa}$ is negative  due to the unphysical
RIS-subtracted         collision        term         $\gamma^{\,\prime
L\Phi}_{\,L^C\Phi^\dagger}$  in~(\ref{GDL2}).  Nevertheless, it can be
shown  that   the   sum    $\Gamma^{\Delta  L   =2}_{\rm  Yukawa}    +
\frac{1}{8}\sum_{i=1,2} \Gamma^{D\;  (i)}$  is  in general  physically
meaningful and   always  positive.  Second, we  have  investigated the
theoretical uncertainties  from  the  IR mass regulator   $m_{\rm IR}$
introduced in Appendix~\ref{app:CT} to deal with IR singularities that
occur in   the  reduced    cross  sections of     certain   reactions.
Specifically,  we  find  that   the  uncertainties  in  the  numerical
predictions are always  less than $10\%$ when  $m_{\rm  IR}$ is varied
from $m_L(T)$  to $m_\Phi  (T)$.  This  last fact provides  additional
confidence on the stability of the numerical results presented here.

\setcounter{equation}{0}
\section{Conclusions}\label{sec:concls}

We  have studied  the scenario  of thermal  leptogenesis in  which the
leptonic asymmetries  are resonantly  amplified through the  mixing of
nearly degenerate heavy Majorana  neutrinos that have mass differences
comparable to their decay widths. We have shown that this particularly
interesting scenario of baryogenesis,  which has been termed here {\em
resonant leptogenesis}, can be  realized with heavy Majorana neutrinos
even as light  as 0.5--1~TeV, in complete accordance  with the current
solar and atmospheric neutrino data.  Models that might predict nearly
degenerate heavy Majorana neutrinos at  the TeV and sub-TeV scales and
lead  to   light-neutrino  mass  matrices   compatible  with  neutrino
oscillation data can be  constructed by means of the Froggatt--Nielsen
mechanism.   Alternatively,  specific  E$_6$  models~\cite{witten}  in
which the  lepton number is approximately violated  may also naturally
realize nearly degenerate heavy Majorana neutrinos of TeV mass.

An  important field-theoretic  issue we  have been  addressing  in the
present article is related to the proper subtraction of RIS's from the
lepton-number-violating  scattering processes.   In order  to identify
the proper RIS contributions, we have examined the analytic properties
of  the  pole  and  residue  structures of  a  resonant  $L$-violating
scattering amplitude.  If the RIS's carry  spin as is the case for the
heavy  Majorana neutrinos, the  effect of  spin de-correlation  in the
squared amplitude should be considered as well. We have shown that the
present method  of extracting the  effective decay amplitude  from the
resonant  part of  a scattering  amplitude is  fully equivalent  to an
earlier  developed   resummation  approach~\cite{APRD}  based   on  an
LSZ-type formalism  for the  unstable particles.  In  addition, within
the context of thermal leptogenesis, i.e.\ with a strongly thermalized
bath, all quantum information related to the preparation or production
mechanism of the initial heavy Majorana-neutrino states gets lost.  In
this case, only the pole  positions and residues of a given scattering
amplitude are invariant under weak-basis transformations.  The present
approach naturally embodies these symmetry properties. Hence, it takes
consistently into account the  phenomena of decoherence in the thermal
bath of  the early Universe,  thereby providing a natural  solution to
the so-called initial-state problem.

Our  predictions for  the  BAU have  been  obtained after  numerically
solving the relevant network of BE's, where all dominant contributions
related  to $1\leftrightarrow  2$ and  $2\leftrightarrow  2$ processes
have been  consistently considered.   In particular, we  have included
the enhanced heavy-neutrino self-energy effects on scatterings as well
as  the   most  important  contributions   at  $T\stackrel{>}{{}_\sim}
m_{N_1}$  that  originate from  gauge-mediated  collision terms.   The
self-energy  effects   on  scatterings  provide  new   sources  of  CP
violation.   As a consequence,  the generated  BAU at  $T\sim m_{N_1}$
could  be larger  even  by a  factor of  3,  with respect  to the  one
predicted  without the  inclusion of  these additional  sources  of CP
violation.  Finally, we should stress again that our improved BE's are
not   only  valid   for   a  quantitative   description  of   resonant
leptogenesis, but  they can be  applied to other  thermal leptogenesis
scenarios as well, including those with hierarchical neutrinos.

The minimal leptogenesis model  under study has the unpleasant feature
that it  does not  provide a testable  candidate for cold  dark matter
(CDM). For  example, the  CDM problem could  be solved  by introducing
additional massive sterile neutrinos  into the theory, but these could
not  be  experimentally observed.   Another  more attractive  solution
would  be   to  consider  supersymmetric  versions  of   the  SM  with
right-handed  neutrinos, where  the  lightest supersymmetric  particle
(LSP), e.g.~the  lightest neutralino, is stable  because of $R$-parity
conservation and so  it may qualify as CDM.   A non-supersymmetric but
equally  appealing  alternative would  be  to  consider the  invisible
Peccei--Quinn   (PQ)   axion~\cite{PQ,ZDFS,KSVZ}  as   CDM~\cite{KT2},
thereby  solving the  known strong  CP  problem on  the same  footing.
Dedicated  experiments~\cite{Sikivie,Zioutas} searching for  PQ axions
can probe this hypothesis.

Although  we  have demonstrated  that  heavy  Majorana neutrinos  with
sub-TeV masses can still be  responsible for the observed BAU, without
being  in conflict  with neutrino  data,  one may  raise the  question
whether this  resonant leptogenesis scenario can give  rise to further
predictions                for                lepton-flavour-violating
processes~\cite{IP,KPS_Z,LFV}, e.g.~for decays $\mu \to e\gamma$, $\mu
\to eee$, $\mu$-$e$ conversion in  nuclei etc.  Here, we should recall
that     the    out-of-equilibrium     constraints     discussed    in
Section~\ref{sec:BEs}.1 imply rather suppressed Yukawa-couplings, thus
leading  to  unobservably  small  lepton-flavour-violating  phenomena.
However, resonant leptogenesis  allows for significant departures from
the   out-of-equilibrium   constraints,    which   in   turn   implies
significantly  less suppressed  Yukawa  couplings.  It  would be  very
interesting to  study in  detail the phenomenological  implications of
this  exciting scenario  of resonant  leptogenesis for  low-energy and
collider experiments.

\subsection*{Acknowledgements}

This work was supported in part by PPARC grant no:~PPA/G/O/2001/00461.
AP  thanks the  organizers of  the workshop  on {\em  Baryogenesis} at
Michigan University (10--28  June 2003), Wilfried Buchm\"uller, Gordon
Kane and Carlos Wagner, for the inspiring atmosphere.

\subsection*{Note added} 

While revising our paper,    we  became aware of~\cite{GNRRS}    where
thermal  effects on  the  collision  terms,   relevant to the   domain
$T\stackrel{>}{{}_\sim} m_{N_1}$, were computed.  Unlike~\cite{GNRRS},
we have included CP-violating contributions to the BE~(\ref{BEL}) from
$2\leftrightarrow  2$  scatterings  which  can  be  several  orders of
magnitude larger than those from   $1\leftrightarrow 2$ decays in  the
above temperature  domain, at least    as is suggested  in the   $T=0$
approximation. In the same context, we have  also considered the extra
$2\leftrightarrow 2$ wash-out terms  which originate from the resonant
exchange of  heavy  Majorana neutrinos in  the  $3\to 2$  scatterings.
Finally,  we reiterate    that resonant leptogenesis   compatible with
neutrino  data  requires sizeable out-of-equilibrium departure factors
$K_i \stackrel{>}{{}_\sim}  30$.   As a  result, $2\leftrightarrow  2$
scatterings  thermalize the plasma   much   faster, so the impact   of
thermal effects on   such leptogenesis scenarios  for  the interesting
region  $T\stackrel{<}{{}_\sim} m_{N_1}$ becomes even less significant
with respect to scenarios with~$K_i \sim 1$.

\newpage

\def\theequation{\Alph{section}.\arabic{equation}}
\begin{appendix}

\setcounter{equation}{0}
\section{Three-heavy-Majorana-neutrino mixing}\label{app:mixing}

It is interesting to see how the analytic expressions for the resummed
$N_iN_j$-propagators $S_{ij}(\not\!  p)$ given in~(\ref{Sresum}) for a
model with   two    heavy Majorana   neutrinos generalize    to    the
three-heavy-neutrino case.

Our  starting     point     is  the  inverse      one-loop   corrected
$N_iN_j$-propagator matrix:
\begin{equation}
 \label{Sinvres} 
S^{-1}_{ij} (\not\!  p)\ =\ \delta_{ij}\, (\not\! p - m_{N_i} )\: 
                                        +\: \Sigma_{ij} (\not\! p )\ =\
\not\!\! D_{ij}(\not\! p)\,,
\end{equation}
where  $i,j=1,2,3$ and  $\not\!\!  D_{ij}(\not\!   p)$  is analogously
defined to the three-generation case~[cf.~(\ref{Dij})].  The inversion
of the  3-by-3 matrix-valued  matrix in~(\ref{Sinvres}) gives  rise to
the resummed  $N_iN_j$-propagator matrix $S_{ij}(\not\!\!   p)$.  ~~In
doing so, it proves useful to introduce first the spinorial quantities
(no sum over repeated indices):
\begin{equation}
  \label{Dijk}
\not\!\!D^{(k)}_{ij}(\not\! p )\ =\ \not\!\!D_{ij}(\not\! p ) \: - \:  
      \not\!\!D_{ik}(\not\! p ) \not\!\!D^{-1}_{kk}(\not\! p ) 
            \not\!\!D_{kj}(\not\! p )\; ,
\end{equation}
where $k\neq i$ and  $k\neq j$, and  $\not\!\!D^{-1}_{kk}(\not\! p ) =
[\not\!\!   D_{kk}(\not\!   p )]^{-1}$.   ~Making use of~(\ref{Dijk}),
the resummed $N_iN_j$-propagators may be expressed as
\begin{eqnarray}
  \label{SN11}
S_{11}(\not\! p) &=& 
\Big( \not\!\!D^{(3)}_{11}\: -\: \not\!\!D^{(3)}_{12}\,
\not\!\!D^{(3)-1}_{22}\not\!\!D^{(3)}_{21}\, \Big)^{-1}\ =\
\Big( \not\!\!D^{(2)}_{11}\: -\: \not\!\!D^{(2)}_{13}\,
\not\!\!D^{(2)-1}_{33}\not\!\!D^{(2)}_{31}\, \Big)^{-1}\,,\\[3mm]
  \label{SN22}
S_{22}(\not\! p) &=& 
\Big( \not\!\!D^{(3)}_{22}\: -\: \not\!\!D^{(3)}_{21}\,
\not\!\!D^{(3)-1}_{11}\not\!\!D^{(3)}_{12}\, \Big)^{-1}\ =\
\Big( \not\!\!D^{(1)}_{22}\: -\: \not\!\!D^{(1)}_{23}\,
\not\!\!D^{(1)-1}_{33}\not\!\!D^{(1)}_{32}\, \Big)^{-1}\,,\\[3mm]
  \label{SN12}
S_{12}(\not\! p) &=& -\, S_{11}(\not\! p)\,
\not\!\!D^{(3)}_{12}\, \not\!\!D^{(3) -1}_{22}\ =\ 
-\, \not\!\!D^{(3) -1}_{11} \not\!\!D^{(3)}_{12}\, S_{22}(\not\!p)\; ,
\end{eqnarray}
where  the dependence of $\not\!\!D_{ij}$  on $\not\!  p$ has not been
displayed.  The remaining entries of $S_{ij}(\not\!  p)$ can easily be
obtained by obvious  cyclic permutations of the indices $i,j,k=1,2,3$.
Exactly as in  the two-generation case, there are  now 3 complex poles
for each  resummed heavy-neutrino  propagator.  For non-trivial mixing
among the  heavy neutrinos,  the  3  complex  pole  positions can   be
calculated by~(\ref{poles}).

The  resummed decay  amplitudes ${\cal  T}  (N_i  \to  L\Phi)$ can  be
calculated by following a line of steps  similar to those presented in
Section~\ref{sec:RIS}.2.   For example, we  find   for ${\cal T}_{N_1}
(N_1\to L\Phi)$
\begin{equation}
  \label{T3gN1}
{\cal T}_{N_1} \ =\ \bar{u}_l\, \Big(\, \Gamma_1\: -\: \Gamma_2\,
\not\!\!D^{(3)-1}_{22}\not\!\! D^{(3)}_{21}\: -\: \Gamma_3\,
\not\!\!D^{(2)-1}_{33}\not\!\! D^{(2)}_{31}\,\Big)\,u_{N_1} (p)\; .
\end{equation}
Here, $\Gamma_{1,2,3}$ contain the proper  vertex corrections to $\Phi
L   N_{1,2,3}$.  The   remaining  resummed  decay  amplitudes   ${\cal
T}_{N_{2,3}}$ exhibit  an  analogous analytic  form. Again,  only  the
absorptive part of the self-energies and vertices  are relevant in the
OS renormalization scheme.

Exactly  as   in     the  two-generation    case,   we    may  compute
from~(\ref{T3gN1})   the  corresponding    resummed  effective  Yukawa
couplings $\bar{h}^\nu_\pm$   in terms   of  self-energy   and  vertex
absorptive   parts   $A_{ij}$  and $B_{li}$,  defined  in~(\ref{Sabs})
and~(\ref{Vabs})   and  appropriately extended   to a three-generation
model.  Neglecting   self-energy   terms that   are  formally   ${\cal
O}[(h^\nu)^4]$  in~(\ref{T3gN1}),  we derive  the   resummed effective
Yukawa couplings:
\begin{eqnarray}
  \label{hres3g}
(\bar{h}^\nu_+ )_{li} \!&=&\! 
h^\nu_{li}\, +\, iB_{li}\: -\: i\, \sum_{j,k=1}^3\,
|\varepsilon_{ijk}|\, h^\nu_{lj}\nonumber\\
&&\hspace{-2cm}\times\,\frac{m_{N_i} ( m_{N_i} A_{ij} + m_{N_j} A_{ji}) 
+ R_{ik} \Big[ m_{N_i} A_{kj} ( m_{N_i} A_{ik} + m_{N_k} A_{ki} )
+ m_{N_j} A_{jk} ( m_{N_i} A_{ki} + m_{N_k} A_{ik} ) \Big]}
{ m^2_{N_i}\, -\, 
m^2_{N_j}\, +\, 2i\,m^2_{N_i} A_{jj} + 2i\,{\rm Im}R_{ik}\,
\Big( m^2_{N_i} |A_{jk}|^2 + 
m_{N_j} m_{N_k} {\rm Re}A^2_{jk}\Big)   }\ ,\nonumber\\
\end{eqnarray}
where
\begin{equation}
R_{ij}\ =\ \frac{m^2_{N_i}}{m^2_{N_i} - m^2_{N_j} + 2i\, m^2_{N_i} 
A_{jj}}
\end{equation}
and $|\varepsilon_{ijk}|$ is   the modulus of the usual   Levi--Civita
anti-symmetric tensor.  The  respective  CP-conjugate effective Yukawa
couplings     $(\bar{h}^\nu_-)_{li}$          are     easily  obtained
from~(\ref{hres3g})    by   replacing the   ordinary  Yukawa couplings
$h^\nu_{li}$ by their complex conjugates.   In the decoupling limit of
$m_{N_3}  \gg   m_{N_{1,2}}$, (\ref{hres3g})  can  be approximated  by
(\ref{hres}).  Using   the     resummed  effective  Yukawa   couplings
$(\bar{h}^\nu_\pm)_{li}$    derived    in    this   appendix,  it   is
straightforward to compute the  leptonic asymmetries $\delta_{N_i}$ in
(\ref{deltaNi}) for the three-generation mixing case.

\newpage

\setcounter{equation}{0}
\section{CP-conserving collision terms}\label{app:CT}

In this appendix we present analytic expressions for the CP-conserving
collision   terms  required   for  the   numerical  solution   of  the
BEs~(\ref{BEN}) and~(\ref{BEL}).

For computational convenience  and comparison with the literature, the
following rescaled variables will be used:
\begin{equation}
z = \frac{m_{N_1}}{T}\,, \qquad x = \frac{s}{m_{N_1}^2}\,, \qquad  
a_i = \left(\frac{m_{N_i}}{m_{N_1}}\right)^2,\qquad 
a_r = \left(\frac{m_{\rm IR}}{m_{N_1}}\right)^2,
\end{equation}
where  $s$ is the   usual Mandelstam variable and   $m_{\rm IR}$ is an
infra-red (IR) mass regulator to be discussed below.

In terms of the  resummed effective Yukawa couplings $\bar{h}^\nu_\pm$
given  in~(\ref{hres})  and~(\ref{hresC}),  the  radiatively corrected
total decay width $\Gamma_{N_i}$ is given by
\begin{equation}
\Gamma_{N_i}\ =\ \frac{m_{N_i}}{16\pi}\
\Big[\, (\bar{h}^{\nu\,\dagger}_+ \bar{h}^\nu_+)_{ii}\: +\: 
(\bar{h}^{\nu\,\dagger}_- \bar{h}^\nu_-)_{ii}\,\Big]\; .
\end{equation}
Using the latter, we may also define the auxiliary parameters $c_i$ as
\begin{equation}
  \label{cj}
c_i\ =\ \left(\,\frac{\Gamma_{N_i}}{m_{N_1}}\,\right)^2\; .
\end{equation}

We  will  now  employ   the  formula~(\ref{gamma})  to  calculate  the
CP-conserving collision terms for $1\to 2$ and $2\to 2$ processes that
occur  in the  BEs~(\ref{BEN})  and~(\ref{BEL}).  These  CP-conserving
collision terms have been  defined as $\gamma^X_Y$ in~(\ref{CT}) for a
generic   process   $X\to   Y$   and  its   CP-conjugate   counterpart
$\overline{X}\to \overline{Y}$.

For   a  $1\to   2$   process,  e.g.\   $N_i\to   L\Phi$  or   $N_i\to
L^C\Phi^\dagger$,   the  corresponding  CP-conserving   collision  term
$\gamma^{N_i}_{L\Phi}$ is found to be
\begin{eqnarray}
\gamma^{N_i}_{L\Phi}\ =\ \gamma (N_i \to L\Phi)\: +\: \gamma (N_i \to
L^C\Phi^\dagger) \!&=&\! \Gamma_{N_i}\, g_{N_i}\, \int \frac{d^3{\bf
p}_{N_i}}{(2\pi)^3}\,\frac{m_{N_i}}{E_{N_i}({\bf p})}\, e^{-E_{N_i}({\bf
p})/T} \nonumber\\
&=&\! \frac{m^4_{N_1} a_i\,\sqrt{c_i}}{\pi^2\, z}\ K_1(z
\sqrt{a_i})\,,
\end{eqnarray}
where $g_{N_i} = 2$ is the number of internal degrees of freedom of $N_i$,
$E_{N_i}({\bf p}) = \sqrt{{\bf p}^2 + m^2_{N_i}}$, and $K_n(z)$ is an
$n$th-order modified Bessel function~\cite{AS}.

For  $2\to   2$  processes,  the  usual  definition   of  the  reduced
cross-section introduced in~\cite{MAL} was used:
\begin{equation}
  \label{reducedxs}
\widehat{\sigma}(s)\ \equiv\ 8\pi\,\Phi (s)\int\! d\pi_Y\: (2\pi)^4
\,\delta^{(4)} (q-p_Y)\: \left|{\cal M}(X\rightarrow Y)\right|^2\; ,
\end{equation}
where  $s =  q^2$ and  $|{\cal M}(X\to  Y)|^2$ is  the  squared matrix
element summed over all internal degrees of freedom of the initial and
final  states.   In~(\ref{reducedxs}),  $\Phi  (s)$ is  the  so-called
initial phase space integral defined as
\begin{equation}
  \label{Phi}
\Phi (s)\ \equiv\ \int\! d\pi_X\: (2\pi)^4\,\delta^{(4)} (p_X-q)\,.
\end{equation}
The  above expressions simplify  to give  the more  practically useful
equation
\begin{equation}
  \label{sigmat}
\widehat{\sigma}(s)\ =\ \frac{1}{8\pi s}\
\int\limits_{t_-}^{t_+}\! dt\ \left|{\cal M}(X\rightarrow Y)\right|^2\; ,
\end{equation}
where $t$ is the usual Mandelstam variable.

In the calculation of the  reduced cross-sections, we face the problem
that not all of them are IR safe.  In processes, such as $N_i V_\mu\to
L\Phi$, the exchanged  particles, e.g.\ $\Phi$ and  $L$, that occur in
the $t$ and $u$ channels are massless.  Thus, one finds divergences at
the $t$-integration  limits  $t_\pm$  when performing  the phase-space
integral in~(\ref{sigmat}).  A more appropriate framework to deal with
this   problem  is finite-temperature   field  theory,  where such  IR
singularities  are  regulated by the thermal   masses of the particles
involved  in the reaction.   In  fact, although  thermal  mass effects
break the manifest Lorentz invariance, they preserve chirality and the
gauge  symmetries of  the   theory~\cite{Bellac}.  In our  $T=0$ field
theory calculation of the collision   terms, we have regulated the  IR
divergences by cutting off the phase-space integration limits $t_\pm$,
using  a  universal  IR-mass  regulator $m_{\rm   IR}$  related to the
thermal  masses of    the exchanged  particles.     Evidently, our  IR
regularization preserves chirality    and gauge   invariance,  as   is
expected from   a finite-$T$ calculation.    More explicitly,  for the
reduced  cross-sections  (\ref{NutoLQ}),       (\ref{NLtoHV})      and
(\ref{NHtoLV}) which  are  calculated below,  the following upper  and
lower limits of $t$ were used:
\begin{equation}
t_+\ =\ -\,m_{\rm IR}^2\,,\qquad t_-\ =\ m_{N_i}^2\, -\, s\; .
\end{equation}
The reduced cross-section~(\ref{NVtoLH}) was computed by 
cutting-off the integral at both $t$-limits, with
\begin{equation}
t_+\ =\ -\,m_{\rm IR}^2\,,\qquad t_-\ =\ m_{N_i}^2\, +\, m_{\rm
IR}^2\, -\, s\; .
\end{equation}
The IR-mass  regulator  $m_{\rm  IR}$  is chosen to  vary  between the
lepton  and Higgs  thermal  masses, $m_L(T)$   and  $m_\Phi (T)$,   at
$T\approx m_{N_1}$ [cf.~(\ref{thermal})].  The resulting variations in
the  predictions of  the  number  densities  should be  considered  as
theoretical uncertainties due to our $T=0$ field-theory calculation of
the collision terms.

A simple formula  can be found for  the  CP-conserving collision terms
pertinent to  $2\to 2$ processes, if (\ref{reducedxs}) and~(\ref{Phi})
are   inserted  into (\ref{gamma}).     In this   way,  we obtain  the
expression,
\begin{equation}
  \label{22CT}
\gamma^X_Y\ =\ \frac{m^4_{N_1}}{64\,\pi^4 z}\
\int\limits_{x_{\rm thr} }^\infty\! dx\ 
\sqrt{x}\;K_1(z\sqrt{x})\;\widehat{\sigma}^X_Y (x)\; ,
\end{equation}
where $x_{\rm  thr}$ is the kinematic  threshold for a  given $2\to 2$
process.   In   addition,  the  CP-conserving   reduced  cross-section
$\widehat{\sigma}^X_Y$  in~(\ref{22CT})   is  defined  analogously  to
$\gamma^X_Y$ in (\ref{CT}), i.e.\
\begin{equation}
  \label{redXY}
\widehat{\sigma}^X_Y\ \equiv\ \widehat{\sigma}(X\to Y)\: +\: 
             \widehat{\sigma}(\overline{X} \to \overline{Y})\ =\
             \widehat{\sigma}^{\,Y}_X \; ,
\end{equation}
where the last equality follows from CPT invariance.

\begin{figure}[t]
\begin{center}
\begin{picture}(470,70)(0,30)
\SetWidth{0.8}
\newsavebox{\diagramNLQu}
\savebox{\diagramNLQu}(150,100)[bl]{

\Text(10,20)[]{$L$}
\Text(10,80)[]{$N_i$}
\Text(140,20)[]{$u^C$}
\Text(140,80)[]{$Q$}
\Text(75,60)[]{$\Phi$}
\ArrowLine(20,20)(50,50)
\Line(20,80)(50,50)
\DashArrowLine(50,50)(100,50){5}
\ArrowLine(130,20)(100,50)
\ArrowLine(100,50)(130,80)
}
\newsavebox{\diagramQta}
\savebox{\diagramQta}(150,100)[bl]{

\Text(10,25)[]{$u^C$}
\Text(10,75)[]{$N_i$}
\Text(140,25)[]{$Q^C$}
\Text(140,75)[]{$L$}
\Text(65,50)[]{$\Phi$}
\ArrowLine(75,25)(22,25)
\Line(22,75)(75,75)
\DashArrowLine(75,25)(75,75){5}
\ArrowLine(127,25)(75,25)
\ArrowLine(75,75)(127,75)
}
\newsavebox{\diagramQtb}
\savebox{\diagramQtb}(150,100)[bl]{

\Text(10,25)[]{$Q$}
\Text(10,75)[]{$N_i$}
\Text(140,25)[]{$u$}
\Text(140,75)[]{$L$}
\Text(65,50)[]{$\Phi$}
\ArrowLine(22,25)(75,25)
\Line(22,75)(75,75)
\DashArrowLine(75,25)(75,75){5}
\ArrowLine(75,25)(127,25)
\ArrowLine(75,75)(127,75)
}
\put(0,0){\usebox{\diagramNLQu}}
\put(160,0){\usebox{\diagramQta}}
\put(320,0){\usebox{\diagramQtb}}
\end{picture}
\end{center}
\bigskip
\caption{\em $\Delta L=1$ interactions between leptons, heavy Majorana
neutrinos and quarks.}\label{fig:rc1}
\end{figure}
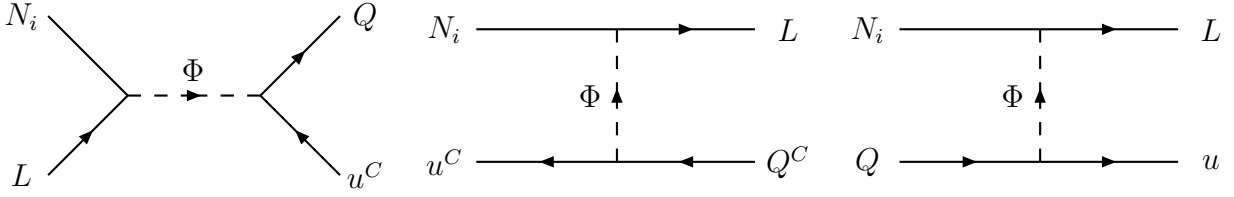

In the following, we present analytic results of CP-conserving reduced
cross-sections for all $2\to 2$  reactions that contribute to the BEs.
As  is  shown  in~\ref{fig:rc1},  we  start  by  listing  the  reduced
cross-sections involving  $\Delta L = 1$  transitions between leptons,
heavy Majorana neutrinos and quarks.  To leading order in $a_r$, these
are given  by~\footnote{Up to an  overall factor, our  analytic result
in~(\ref{NutoLQ})   (specifically  the  coefficient   multiplying  the
non-logarithmic term proportional to $-a_i/x$) agrees with~\cite{MAL},
but   differs  from  the   one  stated   in~\cite{BBP,GCBetal}.   This
discrepancy may be traced in  the different methods used to regularize
the IR singularities.}
\begin{eqnarray}
  \label{NLtoQu}
\widehat{\sigma}^{N_iL}_{Qu^C} & = & 3\,\alpha_u\,
\Big[\, (\bar{h}^{\nu\,\dagger}_+ \bar{h}^\nu_+)_{ii}\: +\: 
(\bar{h}^{\nu\,\dagger}_- \bar{h}^\nu_-)_{ii}\,\Big]\,
\left(\frac{x-a_i}{x}\right)^2\;,\\[3mm]
  \label{NutoLQ}
\widehat{\sigma}^{N_iu^C}_{LQ^C} & = & \widehat{\sigma}^{N_iQ}_{Lu}\nonumber\\ 
&=& 3\,\alpha_u\,
\Big[\, (\bar{h}^{\nu\,\dagger}_+ \bar{h}^\nu_+)_{ii}\: +\: 
(\bar{h}^{\nu\,\dagger}_- \bar{h}^\nu_-)_{ii}\,\Big]\,
\left[1-\frac{a_i}{x}+\frac{a_i}{x}
\ln\left(\frac{x-a_i+a_r}{a_r}\right)\right]\; .
\end{eqnarray}
In~(\ref{NLtoQu}) and (\ref{NutoLQ}), we have defined
$$\alpha_u\ =\   \frac{{\rm   Tr}(h^{u \dagger}  h^u)}{4\pi}\  \simeq\
\frac{\alpha_w\,m_t^2}{2\, M_W^2}\;,$$ where  and $h^u$ is the  up-quark
Yukawa-coupling matrix, and $m_t$ is the top-quark mass.

\begin{figure}[t]
\begin{center}
\begin{picture}(350,180)(0,20)
\newsavebox{\diagramLPLPs}
\savebox{\diagramLPLPs}(150,100)[bl]{
\SetWidth{0.8}

\Text(10,20)[]{$\Phi$}
\Text(10,80)[]{$L$}
\Text(140,20)[]{$L^C$}
\Text(140,80)[]{$\Phi^\dagger$}
\Text(75,60)[]{$N_i$}
\DashArrowLine(20,20)(50,50){5}
\ArrowLine(20,80)(50,50)
\Line(50,50)(100,50)
\ArrowLine(130,20)(100,50)
\DashArrowLine(130,80)(100,50){5}
\Text(75,10)[]{\bf (a)}
}
\newsavebox{\diagramLPLPt}
\savebox{\diagramLPLPt}(150,100)[bl]{
\SetWidth{0.8}

\Text(10,25)[]{$\Phi$}
\Text(10,75)[]{$L$}
\Text(140,25)[]{$L^C$}
\Text(140,75)[]{$\Phi^\dagger$}
\Text(65,50)[]{$N_i$}
\DashArrowLine(22,25)(75,25){5}
\ArrowLine(22,75)(75,75)
\Line(75,25)(75,75)
\ArrowLine(127,25)(75,25)
\DashArrowLine(127,75)(75,75){5}
\Text(75,10)[]{\bf (b)}
}
\newsavebox{\diagramLLPPt}
\savebox{\diagramLLPPt}(150,100)[bl]{
\SetWidth{0.8}

\Text(10,25)[]{$L$}
\Text(10,75)[]{$L$}
\Text(140,25)[]{$\Phi^\dagger$}
\Text(140,75)[]{$\Phi^\dagger$}
\Text(65,50)[]{$N_i$}
\ArrowLine(22,25)(75,25)
\ArrowLine(22,75)(75,75)
\Line(75,25)(75,75)
\DashArrowLine(127,25)(75,25){5}
\DashArrowLine(127,75)(75,75){5}
\Text(75,10)[]{\bf (c)}
}
\newsavebox{\diagramLLPPu}
\savebox{\diagramLLPPu}(150,100)[bl]{
\SetWidth{0.8}

\Text(10,25)[]{$L$}
\Text(10,75)[]{$L$}
\Text(140,25)[]{$\Phi^\dagger$}
\Text(140,75)[]{$\Phi^\dagger$}
\Text(65,50)[]{$N_i$}
\ArrowLine(22,25)(75,25)
\ArrowLine(22,75)(75,75)
\Line(75,25)(75,75)
\DashLine(102.5,50)(75,25){5}
\DashArrowLine(127,75)(102.5,50){5}
\DashLine(75,75)(102.5,50){5}
\DashArrowLine(127,25)(102.5,50){5}
\Text(75,10)[]{\bf (d)}
}
\put(0,100){\usebox{\diagramLPLPs}}
\put(200,100){\usebox{\diagramLPLPt}}
\put(0,0){\usebox{\diagramLLPPt}}
\put(200,0){\usebox{\diagramLLPPu}}
\end{picture}
\end{center}
\bigskip
\caption{\em $\Delta L=2$ interactions between lepton and Higgs
doublets. {\bf (a)} and {\bf (b)} correspond to the process $L\Phi
\leftrightarrow L^C \Phi^\dagger$, {\bf (c)} and {\bf (d)} correspond to the
process $L L \leftrightarrow \Phi^\dagger \Phi^\dagger$.}\label{fig:rc2}
\end{figure}
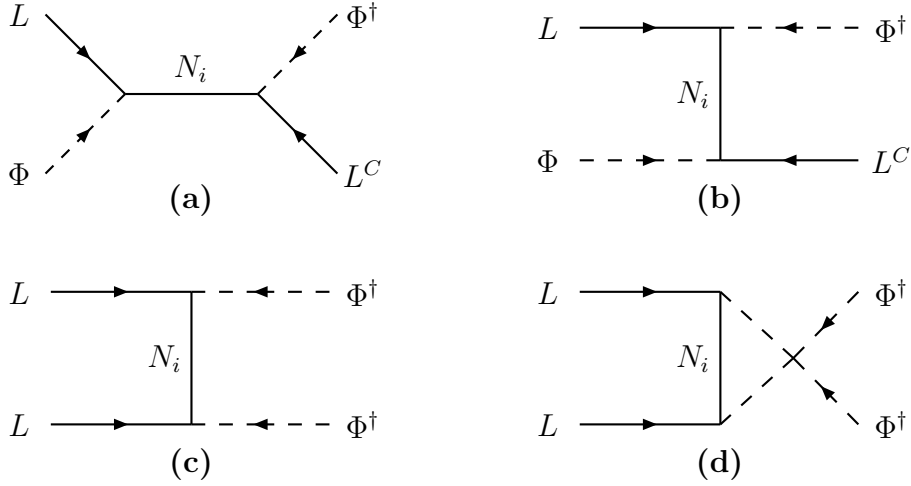

In addition to the above $\Delta L =1$ Higgs-mediated reactions, there
are also $2\leftrightarrow  2$ processes that change the lepton-number
by two units, i.e.~$\Delta L =2$.  As is diagrammatically presented in
Fig~\ref{fig:rc2},  these processes  are: $L\Phi  \leftrightarrow  L^C
\Phi^\dagger$ and $L L \leftrightarrow \Phi^\dagger \Phi^\dagger$, and
their CP-conjugate  counterparts.  For the  first process,  particular
care is needed   to properly  subtract  the  RIS's from  the   reduced
cross-section $\widehat{\sigma}^{\,L\Phi}_{L^C\Phi^\dagger}$.  To this
end, we first define the Breit-Wigner $s$-channel propagators as
\begin{equation}
  \label{Pi}
P^{-1}_i (x) \ =\ \frac{1}{x-a_i+i\sqrt{a_i c_i}}\ .
\end{equation}
Then, the modulus square of a RIS subtracted propagator may be
determined by 
\begin{equation}
  \label{Di}
|D^{-1}_i (x)|^2 \ =\ |P^{-1}_i (x)|^2 \ -\ \frac{\pi}{\sqrt{a_i c_i}}
\; \delta (x - a_i)\ \to \ 0\,.
\end{equation}
This   subtraction  method   is  in   line  with   the  pole-dominance
approximation discussed in  Section~\ref{sec:RIS}, where the last step
of~(\ref{Di})        may       be       obtained        by       means
of~(\ref{Approx})~\footnote{Even   though  our   subtraction  approach
appears to  be similar to the one  suggested recently in~\cite{GNRRS},
our  subtracted  RIS propagator  squared  $|D^{-1}_i (x)|^2$  actually
differs  from~\cite{GNRRS1}  by the  fact  that  we  use the  physical
spectral   representation  of~(\ref{Approx})   for   the  distribution
function  $\delta  (x  -a_i)$,  instead  of  an  arbitrary  regulating
function with the same mathematical features. We have checked that the
difference  of  the   two  approaches  is  numerically  insignificant.
Therefore, in agreement  with earlier remarks in~\cite{APRDremark} and
the  recent observation made  in~\cite{GNRRS}, we  also find  that the
earlier RIS-subtraction approaches, see e.g.~\cite{MAL,BBP,BCST}, tend
to   approximately  overestimate   the  CP-conserving   wash-out  term
$\gamma^{N}_{L\Phi}$ in the BE~(\ref{BEL}) by a factor of~3/2.}.  Note
that  $|D^{-1}_i   (x)|^2$  only  occurs  in   the  squared  amplitude
pertaining  to   an  $s$-channel  diagram.   With  the   help  of  the
newly-defined  quantities in~(\ref{Pi})  and~(\ref{Di}),  the properly
RIS-subtracted                  reduced                  cross-section
$\widehat{\sigma}^{\prime\,L\Phi}_{L^C\Phi^\dagger}$  may be expressed
as follows:
\begin{eqnarray}
   \label{LHtoLH}
\widehat{\sigma}^{\,\prime\, L\Phi}_{L^C\Phi^\dagger}
\!\!&=&\!\!  \sum_{i,j=1}^{3}\ {\rm Re}\, \Bigg\{\,
\Big[(\bar{h}^{\nu\,\dagger}_+ \bar{h}^\nu_+)^2_{ij}\: +\: 
(\bar{h}^{\nu\,\dagger}_- \bar{h}^{\nu}_-)^2_{ij}\Big]
\:\mathcal{A}^{(ss)}_{ij}\: + \: 2 (h^{\nu\,\dagger} h^\nu)^2_{ij}\;
\mathcal{A}^{(tt)}_{ij}\nonumber\\*
\!\!&&\!\! +\:2\, \Big[(\bar{h}^{\nu\,\dagger}_+ h^{\nu})^2_{ij}\: +\: 
(\bar{h}^{\nu\,\dagger}_- h^{\nu *})^2_{ij}\Big]
\mathcal{A}^{(st)*}_{ij}\bigg]\,\Bigg\}\;,
\end{eqnarray}
where
\begin{eqnarray}
  \label{calAij}
\mathcal{A}^{(ss)}_{ij} \!\!&=&\!\! \left\{
\begin{array}{cc}
\frac{\displaystyle x a_i}{\displaystyle 4\pi |D^2_i|}\ \to\ 0\,,
                                                   &\quad (i=j)\,,\\[3mm]
\frac{\displaystyle x\sqrt{a_i\,a_j}}{\displaystyle 
4\pi P^*_i P_j}\ ,&\quad (i\neq j)\,,
\end{array} \right. \nonumber\\*[10pt]
\mathcal{A}^{(st)}_{ij} \!\!&=&\!\! 
\frac{\sqrt{a_i\,a_j}}{2\pi P_i}\left[\, 1\ -\ \frac{x+a_j}{x}\,
\ln\bigg(\frac{x+a_j}{a_j}\,\bigg)\right]\,,\nonumber\\*[10pt]
\mathcal{A}^{(tt)}_{ij} \!\!&=&\!\!
\frac{\sqrt{a_i\,a_j}}{2\pi x\:(a_i-a_j)}\left[\,
(x+a_j)\,\ln\bigg(\frac{x+a_j}{a_j}\bigg)\ -\
(x+a_i)\,\ln\bigg(\frac{x+a_i}{a_i}\bigg)\, \right]\,,\qquad (i \neq j)\,,
\nonumber\\*[10pt]
\mathcal{A}^{(tt)}_{ii} \!\!&=&\!\!  
\frac{a_i}{2\pi x}\Bigg[\, \frac{x}{a_i}\ -\
\ln\bigg(\frac{x+a_i}{a_i}\bigg)\, \Bigg]\,.
\end{eqnarray}

\begin{figure}[t]
\begin{center}
\begin{picture}(350,280)(0,20)
\newsavebox{\diagramNVPLt}
\savebox{\diagramNVPLt}(150,100)[bl]{
\SetWidth{0.8}

\Text(10,25)[]{$V_\mu$}
\Text(10,75)[]{$N_i$}
\Text(140,25)[]{$L$}
\Text(140,75)[]{$\Phi$}
\Text(65,50)[]{$L$}
\Photon(22,25)(75,25){2}{5}
\Line(22,75)(75,75)
\ArrowLine(75,75)(75,25)
\ArrowLine(75,25)(127,25)
\DashArrowLine(75,75)(127,75){5}
\Text(75,10)[]{\bf (a)}
}
\newsavebox{\diagramNVPLu}
\savebox{\diagramNVPLu}(150,100)[bl]{
\SetWidth{0.8}

\Text(10,25)[]{$V_\mu$}
\Text(10,75)[]{$N_i$}
\Text(140,25)[]{$L$}
\Text(140,75)[]{$\Phi$}
\Text(65,50)[]{$\Phi$}
\Photon(22,25)(75,75){2}{5}
\Line(22,75)(75,25)
\DashArrowLine(75,25)(75,75){5}
\ArrowLine(75,25)(127,25)
\DashArrowLine(75,75)(127,75){5}
\Text(75,10)[]{\bf (b)}
}
\newsavebox{\diagramNLPVs}
\savebox{\diagramNLPVs}(150,100)[bl]{
\SetWidth{0.8}

\Text(10,20)[]{$L$}
\Text(10,80)[]{$N_i$}
\Text(140,20)[]{$V_\mu$}
\Text(140,80)[]{$\Phi^\dagger$}
\Text(75,60)[]{$\Phi^\dagger$}
\ArrowLine(20,20)(50,50)
\Line(20,80)(50,50)
\DashArrowLine(100,50)(50,50){5}
\Photon(100,50)(130,20){2}{5}
\DashArrowLine(130,80)(100,50){5}
\Text(75,10)[]{\bf (c)}
}
\newsavebox{\diagramNLPVt}
\savebox{\diagramNLPVt}(150,100)[bl]{
\SetWidth{0.8}

\Text(10,25)[]{$L$}
\Text(10,75)[]{$N_i$}
\Text(140,25)[]{$V_\mu$}
\Text(140,75)[]{$\Phi^\dagger$}
\Text(65,50)[]{$L$}
\ArrowLine(22,25)(75,25)
\Line(22,75)(75,75)
\ArrowLine(75,25)(75,75)
\Photon(75,25)(127,25){2}{5}
\DashArrowLine(127,75)(75,75){5}
\Text(75,10)[]{\bf (d)}
}
\newsavebox{\diagramNPLVs}
\savebox{\diagramNPLVs}(150,100)[bl]{
\SetWidth{0.8}

\Text(10,20)[]{$\Phi^\dagger$}
\Text(10,80)[]{$N_i$}
\Text(140,20)[]{$V_\mu$}
\Text(140,80)[]{$L$}
\Text(75,60)[]{$L$}
\DashArrowLine(50,50)(20,20){5}
\Line(20,80)(50,50)
\ArrowLine(50,50)(100,50)
\Photon(100,50)(130,20){2}{5}
\ArrowLine(100,50)(130,80)
\Text(75,10)[]{\bf (e)}
}
\newsavebox{\diagramNPLVt}
\savebox{\diagramNPLVt}(150,100)[bl]{
\SetWidth{0.8}

\Text(10,25)[]{$\Phi^\dagger$}
\Text(10,75)[]{$N_i$}
\Text(140,25)[]{$V_\mu$}
\Text(140,75)[]{$L$}
\Text(65,50)[]{$\Phi^\dagger$}
\DashArrowLine(75,25)(22,25){5}
\Line(22,75)(75,75)
\DashArrowLine(75,75)(75,25){5}
\Photon(75,25)(127,25){2}{5}
\ArrowLine(75,75)(127,75)
\Text(75,10)[]{\bf (f)}
}
\put(0,200){\usebox{\diagramNVPLt}}
\put(200,200){\usebox{\diagramNVPLu}}
\put(0,100){\usebox{\diagramNLPVs}}
\put(200,100){\usebox{\diagramNLPVt}}
\put(0,0){\usebox{\diagramNPLVs}}
\put(200,0){\usebox{\diagramNPLVt}}
\end{picture}
\end{center}
\medskip
\caption{\em $\Delta L=1$ interactions between heavy Majorana
neutrinos, gauge bosons, lepton doublets and Higgs doublets. {\bf (a)} 
and {\bf (b)} correspond to the process $N_i V_\mu \leftrightarrow \Phi L$,
{\bf (c)} and {\bf (d)} correspond to the process $N_i L \leftrightarrow 
V_\mu \Phi^\dagger$ and {\bf (e)} and {\bf (f)} correspond to the process 
$N_i \Phi^\dagger \leftrightarrow L V_\mu$.}\label{fig:rc3}
\end{figure}
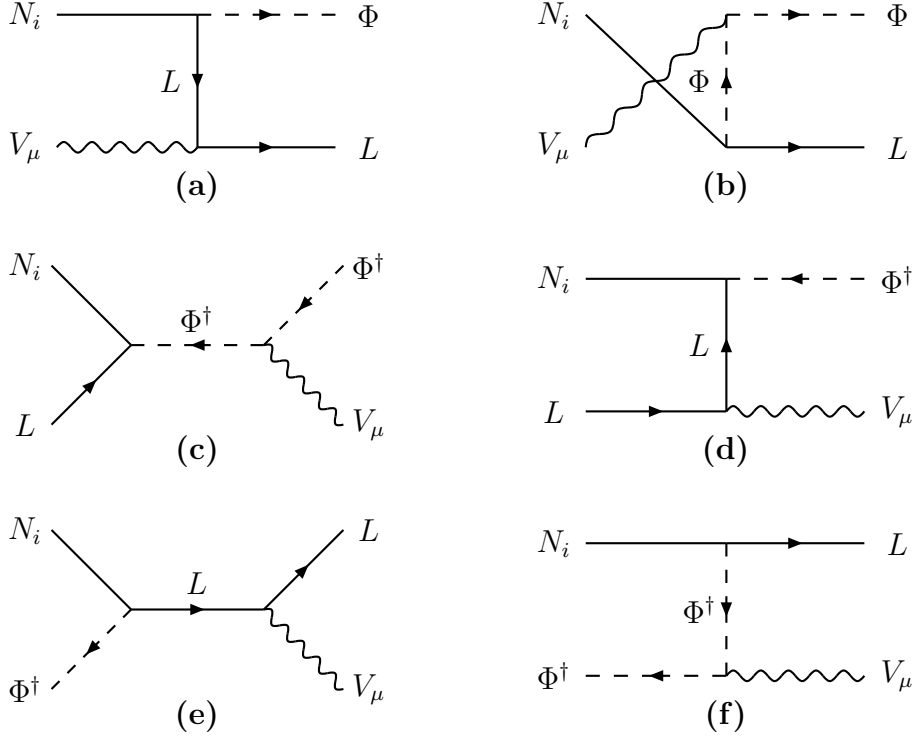

The second $\Delta L=2$ reaction $LL \to \Phi^\dagger\Phi^\dagger$ and
its CP-conjugate one  does not involve RIS's and  hence can be written
down in the shorter form:
\begin{equation}
   \label{LLtoHH}
\widehat{\sigma}^{LL}_{\Phi^\dagger\Phi^\dagger} \ = \
\sum_{i,j=1}^{3}\; {\rm Re}\, \Big[
(h^{\nu \dagger} h^\nu)^2_{ij}\Big]\; \mathcal{B}_{ij}\;,
\end{equation}
where
\begin{equation}
  \label{calBij}
\mathcal{B}_{ij}\ =\ \frac{\sqrt{a_i\,a_j}}{2\pi}
\left[\, \frac{1}{a_i-a_j}
\ln\left(\frac{a_i(x+a_j)}{a_j(x+a_i)}\right)\: +\:
\frac{1}{x+a_i+a_j}
\ln\left(\frac{(x+a_i)(x+a_j)}{a_i\,a_j}\right)\,\right]\,.
\end{equation}
For $i=j$, this last expression simplifies to
\begin{equation}
\mathcal{B}_{ii} \ = \
\frac{1}{2\pi}\left[\,\frac{x}{x+a_i}\: +\: \frac{2\,a_i}{x+2a_i}
\ln\left(\frac{x+a_i}{a_i}\right)\,\right]\, .
\end{equation}
Note that  all the  expressions for ${\cal  A}_{ij}$ in~(\ref{calAij})
and ${\cal B}_{ij}$ in~(\ref{calBij}) vanish individually in the limit
$x \to 0$.

Finally, there are additional $\Delta L =1$ reactions that involve the
SM  gauge bosons  $V_\mu  =     B_\mu,\,W^a_\mu$  and are     depicted
diagrammatically  in Fig.~\ref{fig:rc3}.  To  leading  order in $a_r$,
the corresponding CP-conserving reduced cross-sections are given by
\begin{eqnarray}
  \label{NVtoLH}
\widehat{\sigma}^{N_iV_\mu}_{L\Phi} \!&=&\! 
\frac{n_V g_V^2}{8\pi\,x} \,
\Big[\, (\bar{h}^{\nu\,\dagger}_+ \bar{h}^\nu_+)_{ii}\: +\: 
(\bar{h}^{\nu\,\dagger}_- \bar{h}^\nu_-)_{ii}\,\Big]\,
\left[\frac{(x+a_i)^2}{x-a_i + 2 a_r}\,
\ln\left(\frac{x-a_i + a_r}{a_r}\right)\right]\,,\\[3mm]
  \label{NLtoHV}
\widehat{\sigma}^{N_iL}_{\Phi^\dagger V_\mu} \!&=&\!
\frac{n_V g_V^2}{16\pi\, x^2}\,
\Big[\, (\bar{h}^{\nu\,\dagger}_+ \bar{h}^\nu_+)_{ii}\: +\: 
(\bar{h}^{\nu\,\dagger}_- \bar{h}^\nu_-)_{ii}\,\Big]\nonumber\\
&&\times\, \Bigg[\, (5x-a_i)\,(a_i -x)\: +\: 
2(x^2+xa_i - a^2_i)\,
\ln\left(\frac{x-a_i + a_r}{a_r}\right)\Bigg]\, ,\qquad\quad\\[3mm]
  \label{NHtoLV}
\widehat{\sigma}^{N_i\Phi^\dagger}_{LV_\mu} \!&=&\!
\frac{n_V g_V^2}{16\pi\,x^2} \,
\Big[\, (\bar{h}^{\nu\,\dagger}_+ \bar{h}^\nu_+)_{ii}\: +\: 
(\bar{h}^{\nu\,\dagger}_- \bar{h}^\nu_-)_{ii}\,\Big]\,\nonumber\\
\!&&\! \times\, (x-a_i)\, \Bigg[ x-3a_i\: +\: 4a_i\,
\ln\left(\frac{x-a_i+a_r}{a_r}\right)\Bigg]\, ,
\end{eqnarray}
with  $g_V  =  g',g$  and  $n_V=1,3$ for  $V_\mu  =  B_\mu,\,W^a_\mu$,
respectively.   Notice that  up  to an  overall  factor, the  analytic
expressions for the SU(2)$_L$  and U(1)$_Y$ reduced cross-sections are
identical. The different overall  factors can be evaluated by properly
tracing the gauge degrees of freedom  when a gauge boson couples to an
iso-doublet field  such as $L$  and $\Phi$: $\frac{1}{4}\,  g^2\, {\rm
Tr}\,   (\tau_a   \tau_a)   =   \frac{3}{2}g^2$  for   SU(2)$_L$   and
$\frac{1}{4}\, g'^2\, {\rm Tr}\,  ({\bf 1}_2\,{\bf 1}_2) = \frac{1}{2}
g'^2$ for U(1)$_Y$, where  $\tau_{1,2,3}$ are the usual Pauli matrices
and {\bf 1$_2$} is the $2\times 2$ unit matrix.

\newpage

\end{appendix}


\begin{thebibliography}{9}
\bibitem{APRD1} One may compare this formula with (3.11) in~\cite{APRD}.
\end{thebibliography}
\end{document}